\documentclass[reqno,12pt]{article}
\usepackage{amsmath,amsfonts,amssymb,amsthm,amstext,amscd}
\usepackage{color}
\usepackage[all]{xy}
\usepackage{hyperref}
\usepackage{epsfig}
\usepackage{graphicx}
\usepackage{tikz}
\usetikzlibrary{decorations.pathmorphing}
\usepackage{caption}
\usepackage{subcaption}
\usepackage{lipsum}
\usepackage{psfrag}
\usepackage{epstopdf}
\usepackage{comment}
\usepackage{graphicx}
\usepackage{cite}
\usepackage[utf8,mac]{inputenc}
\makeatletter \@addtoreset{equation}{section}

\makeatletter\renewcommand\section{\@startsection {section}{1}{\z@}%
                                                                                                                                                                                                                                                                                                                                                                                                                                                                                                                                                                                                                                                                                                                                                                                                                                                                                                                                                                                                                                                                                                                                                                                                                                                                                                                                                                                                                                                                                                                                                                                                                                                                                                                                                                                                                                                                                                   {-3.5ex \@plus -1ex \@minus -.2ex}
                                                                                                                                                                                                                                                                                                                                                                                                                                                                                                                                                                                                                                                                                                                                                                                                                                                                                                                                                                                                                                                                                                                                                                                                                                                                                                                                                                                                                                                                                                                                                                                                                                                                                                                                                                                                                                                                                                   {2.3ex \@plus.2ex}%
                                                                                                                                                                                                                                                                                                                                                                                                                                                                                                                                                                                                                                                                                                                                                                                                                                                                                                                                                                                                                                                                                                                                                                                                                                                                                                                                                                                                                                                                                                                                                                                                                                                                                                                                                                                                                                                                                                   {\normalfont\large\bfseries}}
\renewcommand\subsection{\@startsection{subsection}{2}{\z@}%
                                                                                                                                                                                                                                                                                                                                                                                                                                                                                                                                                                                                                                                                                                                                                                                                                                                                                                                                                                                                                                                                                                                                                                                                                                                                                                                                                                                                                                                                                                                                                                                                                                                                                                                                                                                                                                                                                                                                                                                                                                                     {-3.25ex\@plus -1ex \@minus -.2ex}%
                                                                                                                                                                                                                                                                                                                                                                                                                                                                                                                                                                                                                                                                                                                                                                                                                                                                                                                                                                                                                                                                                                                                                                                                                                                                                                                                                                                                                                                                                                                                                                                                                                                                                                                                                                                                                                                                                                                                                                                                                                                     {1.5ex \@plus .2ex}%
                                                                                                                                                                                                                                                                                                                                                                                                                                                                                                                                                                                                                                                                                                                                                                                                                                                                                                                                                                                                                                                                                                                                                                                                                                                                                                                                                                                                                                                                                                                                                                                                                                                                                                                                                                                                                                                                                                                                                                                                                                                     {\normalfont\bfseries}}

\newcommand{\be}{\begin{equation}}
\newcommand{\ee}{\end{equation}}
\newcommand{\bea}{\begin{eqnarray}}
\newcommand{\eea}{\end{eqnarray}}
\newcommand{\bse}{\begin{subequations}}
\newcommand{\ese}{\end{subequations}}
\newcommand{\beqa}{\begin{eqnarray}}
\newcommand{\eeqa}{\end{eqnarray}}
\newcommand{\beqar}{\begin{eqnarray*}}
\newcommand{\eeqar}{\end{eqnarray*}}
\newcommand{\bi}{\begin{itemize}}
\newcommand{\ei}{\end{itemize}}
\newcommand{\bn}{\begin{enumerate}}
\newcommand{\en}{\end{enumerate}}

\newcommand{\ba}{\begin{array}}
\newcommand{\ea}{\end{array}}
\newcommand{\bc}{\begin{center}}
\newcommand{\ec}{\end{center}}
\newcommand{\nnr}{\nonumber \\}
\newcommand{\nn}{\nonumber}

\newcommand{\p}{\partial}

\newcommand{\mn}{{\mu\nu}}

\newcommand{\de}{\delta}

\newcommand{\eps}{\epsilon}
\def\xp{x^+}
\def\xn{x^-}

\newcommand{\bomega}{{\boldsymbol \omega}}
\newcommand{\bTheta}{{\boldsymbol \Theta}}

\newcommand{\bL}{{\boldsymbol L}}
\newcommand{\cL}{{\mathcal L}}

\newcommand{\bE}{{\boldsymbol E}}

\def\half{\frac{1}{2}}

\begin{document}

\begin{titlepage}

\begin{flushright}\vspace{-3cm}
{\small
IPM/P-2015/072 \\
\today }\end{flushright}
\vspace{0.5cm}

\begin{center}
\centerline{{\Large{\bf{Symplectic and Killing Symmetries of AdS$_3$ Gravity:}}}} \vspace{3mm}
\centerline{{\large{\bf{Holographic vs Boundary Gravitons}}}} \vspace{7mm}

\centerline{\large{\bf{G. Comp\`{e}re\footnote{e-mail:gcompere@ulb.ac.be}$^{\dag}$, P. Mao\footnote{e-mail:pujian.mao@ulb.ac.be}$^{\dag}$, A. Seraj\footnote{e-mail:ali\_seraj@ipm.ir}$^{\ddag}$, M.M. Sheikh-Jabbari
\footnote{e-mail:
jabbari@theory.ipm.ac.ir}$^\ddag$}}}

\vspace{5mm}
\normalsize
\bigskip\medskip
$^\dag$ \textit{Universit\'{e} Libre de Bruxelles and International Solvay Institutes
CP 231 B-1050 Brussels, Belgium
}\\
\smallskip
$^\ddag$ \textit{School of Physics, Institute for Research in Fundamental
Sciences (IPM), \\P.O.Box 19395-5531, Tehran, Iran}
\vspace{5mm}

\begin{abstract}
\noindent

The set of solutions to the AdS$_3$ Einstein gravity with Brown-Henneaux boundary conditions is known to be a family of metrics labeled by two arbitrary periodic functions, respectively left and right-moving. It turns out that there exists an appropriate presymplectic form  which vanishes on-shell. This promotes this set of metrics to a phase space in which the Brown-Henneaux asymptotic symmetries become symplectic symmetries in the bulk of spacetime. Moreover, any element in the phase space admits two global Killing vectors. We show that the conserved charges associated with these Killing vectors commute with the Virasoro symplectic symmetry algebra, extending the Virasoro symmetry algebra with two $U(1)$ generators. We discuss that any element in the phase space falls into the coadjoint orbits of the Virasoro algebras and that each orbit is labeled by the $U(1)$ Killing charges. Upon setting the right-moving function to zero and restricting the choice of orbits, one can take a near-horizon decoupling limit which preserves a chiral half of the symplectic symmetries. Here we show two distinct but equivalent ways in which the chiral Virasoro symplectic symmetries in the near-horizon geometry can be obtained as a limit of the bulk symplectic symmetries.

\end{abstract}


\end{center}

\end{titlepage}
\setcounter{footnote}{0}
\renewcommand{\baselinestretch}{1.05}  

\addtocontents{toc}{\protect\setcounter{tocdepth}{2}}
\tableofcontents

\newpage

\section{Introduction and outline}

Although it admits no propagating degrees of freedom (``bulk gravitons''),  three dimensional Einstein gravity is known to admit black holes \cite{Banados:1992wn, Banados:1992gq}, particles \cite{Deser:1983tn, Deser:1983nh}, wormholes  \cite{Brill:1995jv,Brill:1998pr,Skenderis:2009ju} and boundary dynamics \cite{Brown:1986nw,Ashtekar:1996cd,Barnich:2006av}.   Moreover, it can arise as a consistent subsector of higher dimensional matter-gravity theories, see e.g. \cite{Aharony:1999ti, SheikhJabbaria:2011gc}.  Therefore, three-dimensional gravity in the last three decades has been viewed as a simplified and fruitful setup to analyze and address issues related to the physics of black holes and quantum gravity.

In three dimensions the Riemann tensor is completely specified in terms of the Ricci tensor, except at possible defects, and hence all Einstein solutions with generic cosmological constant are locally maximally symmetric. The fact that AdS$_3$ Einstein gravity can still have a nontrivial dynamical content was first discussed in the seminal work of Brown and Henneaux \cite{Brown:1986ed,Brown:1986nw}. There, it was pointed out that one may associate nontrivial conserved  charges, defined at the AdS$_3$ boundary, to diffeomorphisms which preserve prescribed (Brown-Henneaux) boundary conditions. These diffeomorphisms and the corresponding surface charges obey two copies of the Virasoro algebra and the related bracket structure  may be viewed as a Dirac bracket defining (or arising from) a symplectic structure  for these ``boundary degrees of freedom'' or ``boundary gravitons''. It was realized that the Virasoro algebra should be interpreted in terms of a holographic dictionary with a conformal field theory \cite{Strominger:1997eq}. These ideas found a more precise and explicit formulation within the celebrated AdS$_3$/CFT$_2$ dualities in string theory \cite{Kraus:2006wn}. Many other important results in this context have been obtained  \cite{Maldacena:1998bw,Skenderis:2002wp,Maloney:2007ud,Witten:2007kt,Banados:1998gg,Rooman:2000ei,Ryu:2006ef,Balasubramanian:2009bg,SheikhJabbaria:2011gc,Barnich:2012aw,Li:2013pra,Sheikh-Jabbari:2014nya, Maloney:2015ina, Kim:2015qoa}.

{ Recently in \cite{Compere:2014cna} it was shown that the asymptotic symmetries of dS$_3$ with Dirichlet boundary conditions defined as an analytic continuation of the Brown-Henneaux symmetries to the case of positive cosmological constant \cite{Strominger:2001pn} can be defined everywhere into the bulk spacetime. A similar result is expected to follow for AdS$_3$ geometries by analytical continuation, however, few details were given in \cite{Compere:2014cna} (see also \cite{Mitchell:2015ora,Mitchell:2015xga} for related observations).}
In this work, we revisit the Brown-Henneaux analysis {from the first principles} and show that the surface charges and the associated algebra and dynamics can be defined not only on the circle at spatial infinity, but also on any circle inside of the bulk obtained by a smooth deformation which does not cross any geometric defect or topological obstruction.
This result is consistent with the expectation that if a dual 2d CFT exists, it is not only ``defined at the boundary'', but it is defined in a larger sense from the AdS bulk.

Our derivation starts with the set of Ba\~nados geometries \cite{Banados:1998gg} which constitute all locally AdS$_3$ geometries with Brown-Henneaux boundary conditions. We show that the invariant presymplectic form \cite{Barnich:2007bf} (but not the Lee-Wald presymplectic form \cite{Lee:1990nz}) vanishes in the entire bulk spacetime. The charges defined from the presymplectic form  are {hence} conserved everywhere, i.e. they define sympletic symmetries, and they obey an algebra through a Dirac bracket, which is isomorphic to two copies of the Virasoro algebra. In turn, this Dirac bracket defines a lower dimensional non-trivial symplectic form, {the Kirillov-Kostant symplectic form for coadjoint orbits of the Virasoro group \cite{Witten:1987ty}.} In that sense the boundary gravitons may be viewed as \emph{holographic gravitons}: they define a lower dimensional dynamics inside of the bulk. Similar features were {also observed in the near-horizon region of extremal black holes} \cite{Compere:2015bca, Compere:2015mza}.

Furthermore, we will study in more detail the extremal  sector of the phase space. Boundary conditions are known in the decoupled near-horizon region of the extremal BTZ black hole which admit a chiral copy of the Virasoro algebra \cite{Balasubramanian:2009bg}. Here, we extend the notion of decoupling limit to more general extremal metrics in the Ba\~nados family and show that one can obtain {this} (chiral) Virasoro algebra as a limit of the bulk symplectic symmetries, which are defined from the asymptotic AdS$_3$ region all the way to the near-horizon region. We discuss two distinct ways to take the near-horizon limit: at finite coordinate radius (in Fefferman-Graham coordinates) and at wiggling coordinate radius (in Gaussian null coordinates), depending upon the holographic graviton profile at the horizon. We will show that these two coordinate systems lead to the same conserved charges and are therefore equivalent up to a gauge choice. {Quite interestingly, the vector fields defining the Virasoro symmetries take a qualitatively different form in both coordinate systems  which are also distinct from all previous ansatzes for near-horizon symmetries \cite{Carlip:1998uc,Guica:2008mu,Balasubramanian:2009bg,Compere:2015bca, Compere:2014cna,Compere:2015mza}.}

In \cite{Sheikh-Jabbari:2014nya} it was noted that Ba\~nados geometries in general have (at least) two global $U(1)$ Killing vectors (defined over the whole range of the Ba\~nados coordinate system). We will study the conserved charges  $J_\pm$ associated with these two Killing vectors. We will show that these charges commute with the surface charges associated with symplectic symmetries (the Virasoro generators). We then discuss how the elements of the phase space may be labeled using the $J_\pm$ charges. This naturally brings us to the question of how the holographic gravitons may be labeled through representations of Virasoro group, the Virasoro coadjoint orbits, e.g. see \cite{Balog:1997zz, Witten:1987ty}. The existence of Killing horizons in the set of Ba\~nados geometries was studied in \cite{Sheikh-Jabbari:2014nya}.  We discuss briefly that if the Killing horizon exists, its area defines an entropy which together with $J_\pm$, satisfies the first law of thermodynamics.

The organization of this paper is as follows. In section \ref{symp-sym-sec}, we establish that the family of locally AdS$_3$ geometries with Brown-Henneaux boundary conditions forms a phase space with two copies of the Virasoro algebra as symplectic symmetries.  In section \ref{Killing-sec}, we show that each metric in the phase space admits two $U(1)$ Killing vectors which commute with the vector fields generating the symplectic symmetries, once we use the appropriately ``adjusted (Lie) bracket'' \cite{Barnich:2010eb, 2010AIPC.1307....7B}. We show that the charge associated with these two Killing vectors are integrable over the phase space and commute with the generators of the Virasoro symplectic symmetries. In section \ref{Banados-Orbits-sec}, we discuss how the phase space falls into Virasoro coadjoint orbits and how the Killing charges may be attributed to each orbit. We also discuss the first law of thermodynamics on the black hole orbits. In section \ref{extremal-NH-sec}, we focus on a chiral half of the phase space which is obtained through decoupling limit over the extremal geometries. We show that this sector constitutes a phase space with symplectic symmetries of its own.
We discuss the limit in both Fefferman-Graham and Gaussian null coordinate systems. In section \ref{discussion-sec}, we summarize our results and present our outlook.
In appendix \ref{appendix-A-sec}, we review and discuss the covariant phase space method, especially focusing on the case where the vector fields generating the symmetries are field dependent. We present in detail the definition of the surface charges and their integrability condition.

\section{Symplectic symmetries in the bulk spacetime}\label{symp-sym-sec}

The AdS$_3$ Einstein  gravity is described by the action and equations of motion,
\be\label{action-eom}
S=\frac{1}{16\pi G}\int d^3x \sqrt{-g} (R+\frac{2}{\ell^2} ),\qquad R_{\mu\nu}=-\frac{2}{\ell^2}g_{\mu\nu}.
\ee
As discussed in the introduction, all solutions are locally AdS$_3$ with radius $\ell$.
To represent the set of these solutions, we adopt the Fefferman-Graham coordinate system\footnote{We will purposely avoid to use the terminology of Fefferman-Graham \emph{gauge} which would otherwise presume that  leaving the coordinate system by any infinitesimal diffeomorphism would be physically equivalent in the sense that the associated canonical generators to this diffeomorphism would admit zero Dirac brackets with all other physical generators. Since this coordinate choice precedes the definitions of boundary conditions, and therefore the definition of canonical charges, the \emph{gauge} terminology is not appropriate.} \cite{FGpaper:1985fg,Skenderis:2002wp,  Papadimitriou:2005ii},
\begin{align}\label{F-G gauge}
	g_{rr}=\dfrac{\ell^2}{r^2},\qquad g_{ r a}=0,\qquad a=1,2,
\end{align}
where the metric reads
\begin{align}
	ds^2&=\ell^2\dfrac{dr^2}{r^2}+\gamma_{ab}(r,x^c)\,dx^a\,dx^b.
\end{align}
Being asymptotically locally AdS$_3$, close to the boundary $r \rightarrow \infty$ one has the expansion $\gamma_{ab} = r^2 g^{(0)}_{ab}(x^c)+{\cal O}(r^0)$ \cite{Skenderis:2002wp}. A variational principle is then defined for a subset of these solutions which are constrained by a boundary condition. Dirichlet boundary conditions amount to fixing the boundary metric $g^{(0)}_{ab}$. The Brown-Henneaux boundary conditions \cite{Brown:1986nw} are Dirichlet boundary conditions with a fixed flat boundary metric,
\bea\label{BCBH}
g^{(0)}_{ab}dx^a dx^b = -dx^+ dx^-,
\eea
together with the periodic identifications $(x^+,x^-) \sim (x^++2\pi ,x^--2\pi)$ which identify the boundary metric with a flat cylinder (the identification reads $\phi \sim \phi +2 \pi$ upon defining $x^\pm = t/\ell \pm \phi$). Other relevant Dirichlet boundary conditions include the flat boundary metric with no identification (the resulting solutions are usually called ``Asymptotically Poincar\'e AdS$_3$''), and the flat boundary metric with null orbifold identification $(x^+,x^-) \sim (x^++2\pi,x^-)$ which is relevant to describing near-horizon geometries \cite{Coussaert:1994tu,Balasubramanian:2009bg, Sheikh-Jabbari:2014nya}.\footnote{Other boundary conditions which lead to different symmetries were discussed in \cite{Compere:2008us,Compere:2013bya,Troessaert:2013fma}.}

The set of all solutions to AdS$_3$ Einstein gravity with {flat boundary metric} 
was given by Ba\~nados \cite{Banados:1998gg} in the Fefferman-Graham coordinate system. The metric takes the form
\begin{align}\label{Banados}
	ds^2&=\ell^2\dfrac{dr^2}{r^2}-\Big(rdx^+-\ell^2\dfrac{L_-(x^-)dx^-}{r}\Big)\Big(rdx^--\ell^2\dfrac{L_+(x^+)dx^+}{r}\Big)
\end{align}
where $L_\pm$ are two single-valued arbitrary functions of their argument. The determinant of the metric is $\sqrt{-g}=\frac{\ell}{2r^3}(r^4-\ell^4 L_+ L_-)$ and the coordinate patch covers the radial range $r^4 > \ell^4 L_+ L_-$. These coordinates are particularly useful in stating the universal sector of all AdS$_3$/CFT$_2$  correspondences since the expectation values of holomorphic and anti-holomorphic components of the energy-momentum tensor of the CFT can be directly related to $L_\pm$ \cite{Brown:1986nw,Maldacena:1998bw}.

The constant $L_\pm$ cases correspond to better known geometries \cite{Deser:1983nh,Banados:1992wn,Banados:1992gq}: $L_+=L_-=-1/4$ corresponds to AdS$_3$ in global coordinates, $-1/4< L_\pm< 0$ correspond to conical defects (particles on AdS$_3$), $L_-=L_+=0$ correspond to massless BTZ and generic positive values of $L_\pm$ correspond to generic BTZ geometry of mass and angular momentum respectively equal to $(L_++L_-)/(4G)$ and $\ell(L_+-L_-)/(4G)$.  The selfdual orbifold of AdS$_3$ \cite{Coussaert:1994tu} belongs to the phase space with null orbifold identification and $L_-=0, L_+\neq 0$.

\subsection{Phase space in Fefferman-Graham coordinates}\label{Banados-gauge-sec}

We would now like to establish that the set of Ba\~nados metrics \eqref{Banados} together with a choice of periodic identifications of $x^\pm$ forms a well-defined on-shell phase space. To this end, we need to take two steps: specify  the elements in the tangent space of the on-shell phase space and then define the presymplectic structure over this phase space. Given that the set of all solutions are of the form \eqref{Banados}, the on-shell tangent space is clearly given by metric variations {of the form} 
\be\label{generic-perturbations}
\delta g= g(L+\delta L)- g(L)\,,
\ee
where $\delta L_\pm$ are arbitrary single-valued  functions.
The vector space of all on-shell perturbations $\delta g$ {can be written as} the direct sum of two types of perturbations: those which are generated by diffeomorphisms and those which are not, and that we will refer to as \textit{parametric perturbations}.

As for the presymplectic form, there are two known definitions for Einstein gravity: the one $\boldsymbol \omega^{LW}$ by Lee-Wald \cite{Lee:1990nz} (see also Crnkovic and Witten \cite{Crnkovic:1986ex}) and invariant presymplectic form $\boldsymbol \omega^{inv}$ as defined in \cite{Barnich:2007bf}.\footnote{More precisely, $\boldsymbol \omega$ is a $(2;2)$ form i.e. a two-form on the manifold and a two-form in field space. For short we call $\boldsymbol \omega$ a presymplectic form, and given any spacelike surface $\Sigma$, $\Omega=\int_{\Sigma}\boldsymbol \omega$ the associated presymplectic structure, which is the (possibly degenerate)  $(0;2)$ form. A non-degenerate $(0;2)$ form defines a symplectic structure. We also refer to $\boldsymbol \omega$ as a \emph{bulk} presymplectic form.} The invariant presymplectic form is determined from Einstein's equations only, while the Lee-Wald presymplectic form is determined from the Einstein-Hilbert action, see \cite{Compere:2007az} for details. Upon explicit evaluation, we obtain that the invariant presympletic form  exactly vanishes \emph{on-shell} on the phase space determined by the set of metrics \eqref{Banados}, that is,
\bea\label{noo}
\boldsymbol\omega^{inv}[\delta g , \delta g ; g] \approx 0 .
\eea
On the contrary, the Lee-Wald presymplectic form  is equal to a boundary term
\bea\label{noo2}
\boldsymbol \omega^{LW}[\delta g , \delta g ; g] \approx  -d \boldsymbol E[\delta g ,\delta g ; g] ,\qquad \star \boldsymbol E[\delta g ,\delta g ; g]  = \frac{1}{32 \pi G} \delta g_{\mu \alpha} g^{\alpha\beta}\delta g_{\nu \beta} dx^\mu \wedge dx^\nu
\eea
Indeed, the two presymplectic forms are precisely related by this boundary term \cite{Barnich:2007bf}, as reviewed in appendix.  The fact that the invariant presymplectic form  vanishes on-shell illustrates the fact that there are no propagating bulk degrees of freedom in three dimensional vacuum Einstein gravity. Nevertheless, this result does not imply the absence of dynamics. In fact, there is a non-trivial \emph{lower} dimensional dynamics, i.e. there exists surface charges with  non-trivial Dirac brackets, which can also be inverted into a non-trivial spacetime codimension two (i.e. here, dimension one)  presymplectic structure \cite{Witten:1987ty}\footnote{{This fact is also related to the existence of a non-trivial bulk presymplectic form \emph{off-shell} through the fundamental ``generalized Noether theorem for gauge theories'' applied for gravity \cite{Barnich:1995ap,Barnich:2000zw}. Indeed, to any bulk presymplectic form and any vector field $\xi$ one associates a particular \emph{spacetime codimension two-form} and one-form in field space, i.e. a $(1;1)$ form $k_\xi$  which is conserved when $\xi$ is an isometry. This form in turns allows to define a Dirac bracket and a presymplectic structure of spacetime codimension two as well. Usually in field theories, a symplectic structure is defined from the bulk presymplectic structure as a restriction on the part of field space where it is non-degenerate. For gauge theories however and in the absence of bulk local dynamics, the symplectic structure turns out to be defined from a lower dimensional dynamics somehow hidden in the off-shell part (and not the on-shell part) of the bulk presymplectic structure.}}. We avoid here the terminology of boundary dynamics since, as we will discuss below, the point is precisely that the lower dimensional dynamics exists everywhere in the bulk. We hence {prefer to use the terminology} of \emph{holographic gravitons} instead of boundary gravitons. The existence of a lower dimensional dynamics reveals a deep fact about the holographic nature of gravity.

\subsection{Symplectic symmetries and charges}

As mentioned earlier, the most general form of on-shell perturbations preserving Fefferman-Graham coordinates is of the form \eqref{generic-perturbations}. Among them there are perturbations generated by an infinitesimal diffeomorphism along a vector field $\chi$. The components of such vector field are of the form \cite{Barnich:2010eb,Compere:2008us}
\begin{align}\label{Allowed}
	\chi^r=r \,\sigma(x^a),\qquad
	\chi^a=\eps^a(x^b)-\ell^2\p_b \,\sigma \int_r^\infty \dfrac{dr'}{r'} \gamma^{ab}(r',x^a)
\end{align}
where $\sigma(x^a)$ and $\eps^a(x^b)$ are constrained by the requirement $\de g^{(0)}_{ab}\equiv\mathcal{L}_{\vec{\eps}}\,g^{(0)}_{ab}+2\sigma g^{(0)}_{ab}=0$.
That is, $\vec{\eps} \equiv (\eps_+(x^+),\eps_-(x^-))$ is restricted to be a conformal Killing vector of the flat boundary metric and $\sigma$ is defined as the Weyl factor  in terms of $\vec{\eps}$.

One can in fact explicitly perform  the above integral for a given Ba\~nados metric and solve for $\sigma(x)$ to arrive at
\begin{align}\label{symplectic symmetries}
	\chi&=-\dfrac{r}{2}(\eps_+'+\eps_-')\p_r + \Big(\eps_+ +\dfrac{\ell^2r^2 \eps_-''+\ell^4L_- \eps_+''}{2(r^4-\ell^4L_+L_-)}\Big)\p_+ +\Big(\eps_- +\dfrac{\ell^2r^2 \eps_+''+\ell^4L_+ \eps_-''}{2(r^4-\ell^4L_+L_-)}\Big)\p_-,
\end{align}
where $\eps_\pm$ are two arbitrary single-valued periodic functions of $x^\pm$ and possibly of the fields $L_+(x^+),\, L_-(x^-)$, and the \emph{prime} denotes derivative w.r.t. the argument. As we see,
\begin{enumerate}
	\item $\chi$ is a \emph{field-dependent} vector field. That is, even if the two arbitrary functions $\eps_\pm$ are field independent, it has explicit dependence upon $L_\pm$: $\chi=\chi(\eps_\pm; L_\pm)$.
	\item The vector field $\chi$ is defined in the entire coordinate patch spanned by the Ba\~nados metric, not only asymptotically.
	\item Close to the boundary, at large $r$, $\chi$ reduces to the Brown-Henneaux asymptotic symmetries \cite{Brown:1986nw}. Also, importantly, at large $r$ the field-dependence of $\chi$ drops out if one also takes $\eps_\pm$ field-independent.
\end{enumerate}
The above points bring interesting conceptual and technical subtleties, as compared with the better known Brown-Henneaux case, that we will fully address.

The above vector field can be used to define a class of on-shell perturbations, { $\delta_\chi g_{\mu\nu} \equiv\mathcal{L}_{\chi}g_{\mn}$. It can be shown that}
\begin{align}
	\delta_\chi g_{\mn}&=g_{\mn}(L_++\de_\chi L_+, L_-+\de_\chi L_-)-g_{\mn}(L_+,L_-),
\end{align}
where
\be\begin{split}\label{delta-chi-L}
	\de_\chi L_+&= \eps_+ \p_+ L_+ +2L_+ \p_+ \eps_+ -\dfrac{1}{2}\p_+^3\eps_+,\cr
	\de_\chi L_-&= \eps_- \p_- L_- +2L_- \p_- \eps_- -\dfrac{1}{2}\p_-^3\eps_-.
\end{split}\ee
As it is well-known and in the context of AdS$_3$/CFT$_2$ correspondence \cite{Aharony:1999ti, Kraus:2006wn} the variation of $L_\pm$ under  diffeomorphisms generated by $\chi$ is the same as the variation of a 2d CFT energy-momentum tensor under generic infinitesimal conformal transformations. Notably, the last term related to the central extension of the Virasoro algebra is a quantum anomalous effect in a 2d CFT while in the dual  AdS$_3$ gravity it appears  classically.

The vector field $\chi$ determines \emph{symplectic symmetries} as defined in \cite{Compere:2015bca} (they were defined as asymptotic symmetries everywhere in \cite{Compere:2014cna}). The reason is that the invariant presymplectic form contracted with the Lie derivative of the metric with respect to the vector vanishes on-shell,
\begin{align}\label{symplectic condition}
	\boldsymbol \omega^{inv} [ g;\de g,\mathcal{L}_{\chi} g ] & \approx 0,
\end{align}
which is obviously a direct consequence of \eqref{noo}, while $\mathcal{L}_{\chi} g $ does not vanish. Then according to \eqref{charge difference}, the charges associated to symplectic symmetries can be defined over any closed codimension two surface $\mathcal{S}$ (circles in 3d) anywhere in the bulk. Moreover, as we will show next, the surface charge associated to $\chi$ is non-vanishing and integrable. That is, the concept of ``symplectic symmetry'' extends the notion of ``asymptotic symmetry'' inside the bulk.

A direct computation gives the formula for the infinitesimal charge one-forms as defined by Barnich-Brandt \cite{Barnich:2001jy}, see appendix, as
\bea
\boldsymbol k^{BB}_\chi [\delta g ; g]= \boldsymbol{\hat{ k}}_\chi[\delta g ; g] + d \boldsymbol B_\chi[\delta g ; g],
\eea
where
\bea\label{chk}
\boldsymbol{\hat{ k}}_\chi [\delta g ; g]= \frac{\ell}{8\pi G} \left(\eps_+(x^+,L_+(x^+)) \delta L_+ (x^+) dx^+ -  \eps_-(x^-,L_-(x^-)) \delta L_-(x^-) dx^- \right),
\eea
is the expected result and $$\boldsymbol B_\chi =\dfrac{\ell(\eps_+' + \eps_-')(L_+\delta L_- - L_-\delta L_+)}{32\pi G(r^4 - L_+L_-)},$$ is an uninteresting boundary term which vanishes close to the boundary and which drops after integration on a circle.

Now, since the Lee-Wald presymplectic form  does not vanish, the Iyer-Wald \cite{Iyer:1994ys} surface charge one-form is not conserved in the bulk. From the general theory, it differs from the Barnich-Brandt charge by the supplementary term $\boldsymbol E[\delta g ,\mathcal L_\chi g ; g]$, see \eqref{noo2}. In Fefferman-Graham coordinates we have $\mathcal L_\chi g_{r\mu} = 0$ therefore $E_+ = E_- = 0$ and only $E_r$ is non-vanishing. In fact we find $E_r = O(r^{-6})$ which depends upon $L_\pm(x^\pm)$. Since  $\boldsymbol E$ is clearly not a total derivative, the Iyer-Wald charge is explicitly radially dependent which is expected since $\chi$ does not define a symplectic symmetry for the Lee-Wald presymplectic form.

We shall therefore only consider the invariant presymplectic form  and Barnich-Brandt charges here. The standard charges are obtained by considering $\eps_\pm$ to be field-independent. In that case the charges are directly integrable, see also the general analysis of appendix \ref{appendix-A-sec}. We define the left and right-moving stress-tensors as $T=\frac{c}{6} L_+(x^+)$ and $\bar T = \frac{c}{6} L_-(x^-)$ where $c=\frac{3\ell}{2G}$ is the Brown-Henneaux central charge. The finite surface charge one-form then reads
\bea
\boldsymbol Q_\chi [g] \equiv \int^g_{\bar g} \boldsymbol k_\chi [\delta g ; g] = \dfrac{1}{2\pi} (\eps_+(x^+)  T(x^+) dx^+ - \eps_-(x^-) \bar T(x^-) dx^-) .
\eea
Here we chose to normalize the charges to zero for the zero mass BTZ black hole $\bar g$ for which $L_\pm=0$.\footnote{As we will discuss in section \ref{Banados-Orbits-sec}, the zero mass BTZ can only be used as a reference to define charges over a patch of phase space connected to it. For other disconnected patches, one should choose other reference points. } In AdS$_3$/CFT$_2$, the functions $T,\bar T$ are interpreted as components of the dual stress-energy tensor.
In the case of periodic identifications leading to the boundary cylinder (asymptotically global AdS$_3$), we are led to the standard Virasoro charges
\bea\label{ch1}
Q_\chi [g] = \int_S \boldsymbol Q_\chi [g]  = \frac{\ell}{8\pi G}\int_0^{2\pi} d\phi \left(\eps_+(x^+) L_+ (x^+) + \eps_-(x^-) L_-(x^-) \right),
\eea
where $\phi \sim \phi +2 \pi$ labels the periodic circle $S$. The charges are manifestly defined everywhere in the bulk in the range of the Ba\~nados coordinates. 

{Let us finally extend the Ba\~nados geometries beyond the coordinate patch covered by Fefferman-Graham coordinates and comment on the existence of singularities. In the globally asymptotically AdS case, the charges \eqref{ch1} are defined by integration on a circle. Since the charges are conserved in the bulk, one can arbitrarily smoothly deform the integration circle and the charge will keep its value, as long as we do not reach a physical singularity or a topological obstruction. Now, if one could deform the circle in the bulk to a single point, the charge would vanish which would be a contradiction. Therefore, the geometries with non-trivial charges, or ``hair'', are singular in the bulk or contain non-trivial topology which would prevent the circle at infinity to shrink to zero. {In the case of global AdS$_3$ equipped with Virasoro hair, the singularities would be located at defects, where the geometry would not be well-defined.  Such defects are just generalizations of other well known defects.} For example, in the case of conical defects we have an orbifold-type singularity (deficit angle) and for the BTZ black hole, the singularities arise from closed time-like curves (CTC) which are located behind the locus $r=0$ in BTZ coordinates \cite{Banados:1992gq}. Removal of the CTC's creates a topological obstruction which is hidden behind the inner horizon of the BTZ geometry.

\subsection{Charge algebra and adjusted bracket}\label{modified-bracket-sec}
	
The algebra of conserved charges is defined from the Dirac bracket
	\bea
	\{ Q_{\chi_1}, Q_{\chi_2} \} = - \delta_{\chi_1} Q_{\chi_2},
	\eea
where the charges have been defined in appendix.
	
	Let us denote the charge associated with the vector $\chi^+_n = \chi( \eps_+ = e^{i n x^+}, \eps_- = 0)$ by $L_n$ and the charge associated with the vector $\chi^-_n = \chi( \eps_+ = 0 , \eps_- = e^{i n x^-})$ by $\bar L_n$.
	From the definition of charges \eqref{ch1} and the transformation rules \eqref{delta-chi-L}, we directly obtain the charge algebra
	\bea\label{cea}
	\{ L_m , L_n \} &=& (m-n)L_{m+n}+ \frac{c}{12}m^3 \delta_{m+n,0}, \cr
	\{ \bar L_m, L_n \} &=& 0,\\
	\{\bar L_m , \bar L_n \} &=& (m-n) \bar L_{m+n}+ \frac{c}{12}m^3 \delta_{m+n,0}, \nonumber
	\eea
	where
	\be
	c=\frac{3\ell}{2G},
	\ee
	is the Brown-Henneaux central charge. These are the famous two copies of the Virasoro algebra. In the central term there is no contribution proportional to $m$ as a consequence of the choice of normalization of the charges to zero for the massless BTZ black hole.
	
	In fact, the algebra represents, up to a central extension, the algebra of symplectic symmetries. There is however one subtlety. The symplectic symmetry generators $\chi$ are field dependent and hence in computing their bracket we need to ``adjust'' the Lie bracket by subtracting off the terms coming from the variations of fields within the $\chi$ vectors \cite{Barnich:2010eb, 2010AIPC.1307....7B}. Explicitly,
	\begin{align}\label{modified-bracket}
		\big[\chi(\eps_1;L),\chi(\eps_2;L)\big]_*&=\big[\chi(\eps_1;L),\chi(\eps_2;L)\big]_{L.B}-\Big(\de^L_{\eps_1} \chi(\eps_2;L)-\de^L_{\eps_2} \chi(\eps_1;L)\Big),
	\end{align}
	where the variations $\de^L_\eps$ are defined as
	\begin{align}\label{field var}
		\de^L_{\eps_1} \chi(\eps_2;L)&=\de_{\eps_1} L\;\dfrac{\p}{\p L}\chi(\eps_2;L).
	\end{align}
	This is precisely the bracket which lead to the representation of the algebra by conserved charges in the case of field-dependent vector fields. We call $[,]_*$ the \emph{adjusted bracket}. Here the field dependence is stressed by the notation $\chi(\eps;L)$. We also avoided notational clutter by merging the left and right sectors into a compressed notation, $\eps = (\eps_+,\eps_-)$ and $L = (L_+,L_-)$.
		
	Using the adjusted bracket, one can show that symplectic symmetry generators  form a closed algebra
	\begin{align}\label{bracket}
		\big[\chi(\eps_1;L),\chi(\eps_2 ;L)\big]_*&=\chi( \eps_1  \eps_2' - \eps_1'  \eps_2 ;L).
	\end{align}
	Upon expanding in modes $\chi^\pm_n$, one  obtains {two copies of} the Witt algebra
	\bea\label{Witt}
	\big[\chi^+_m,\chi^+_n  \big]_*  &=& (m-n)\chi^+_{m+n}, \cr
	\big[\chi^+_m,\chi^-_n  \big]_* &=& 0, \\
	\big[ \chi^-_m, \chi^-_n \big]_{*}  &=& (m-n) \chi^-_{m+n}, \nonumber
	\eea
	which is then represented by the conserved charges as the centrally extended algebra \eqref{cea}.

\subsection{Finite form of symplectic symmetry transformations}\label{finite-transf-sec}
	
	We discussed in the previous subsections that the phase space of Ba\~nados geometries admits a set of non-trivial tangent perturbations generated by the vector fields $\chi$. Then, there exists finite coordinate transformations (obtained by  ``exponentiating the $\chi$'s'') which map a Ba\~nados metric to another one. That is, there are coordinate transformations
	\be\label{Finite-diff}
	x^\pm \to X^\pm=X^\pm(x^\pm, r)\,,\qquad r\to R=R(x^\pm,r),
	\ee
	with $X^\pm,R$ such that the metric $\tilde g_{\mu\nu}=g_{\alpha\beta}\frac{\partial x^\alpha}{\partial X^\mu} \frac{\partial x^\beta}{\partial X^\nu}$ is a Ba\~nados geometry with appropriately transformed $L_\pm$. Such transformations change the physical charges. They are not gauge transformations but are instead  solution or charge generating transformations.
	
	Here, we use the approach of Rooman-Spindel \cite{Rooman:2000ei}. We start by noting that the technical difficulty in ``exponentiating'' the $\chi$'s arise from the fact that $\chi$'s are field dependent and hence their form  changes as we change the functions $L_\pm$, therefore the method discussed in section 3.3 of \cite{Compere:2015mza} cannot be employed here. However, this feature disappears in the large $r$ regime. Therefore, if we can find the form of \eqref{Finite-diff} at large $r$ we can read how the $L_\pm$ functions of the two transformed metrics should be related. Then, the subleading terms in $r$ are fixed  such that the form of the Ba\~nados metric is preserved. This is guaranteed to work as a consequence of Fefferman-Graham's theorem \cite{FGpaper:1985fg}. From the input of the (flat) boundary metric and first subleading piece (the boundary stress-tensor), one can in principle recontruct the entire metric.
	
	{It can be shown that the finite coordinate transformation preserving \eqref{F-G gauge} is }
	\bea
	&&\xp\to X^+=h_+(\xp) + \dfrac{\ell^2}{2r^2} \frac{h_-''}{h_-'}\frac{h_+'}{h_+}+{\cal O}(r^{-4}),\nn\\
	&&r\to R=\frac{r}{\sqrt{h_+' h_-'}}+{\cal O}(r^{-1}),\\
	&&\xn\to X^-=h_-(x^-)+\dfrac{\ell^2}{2r^2} \frac{h_+''}{h_+'}\frac{h_-'}{h_-}+{\cal O}(r^{-4}),\nn
	\eea
	where $h_\pm(x^\pm +2\pi)=h_\pm(x^\pm) \pm 2\pi$, $h_\pm$ are monotonic ($h_\pm' > 0$) so that the coordinate change is a bijection. At leading order (in $r$), the functions $h_\pm$ parametrize a generic conformal transformation of the boundary metric.
	
	Acting upon the metric by the above transformation one can read how the functions $L_\pm$ transform:
	\bea\label{finite-L-transf}
	L_+(x^+) &\to \tilde L_+={h_+'}^2 L_+ - \dfrac{1}{2} S[h_+;x^+],\\
	L_-(x^-) &\to \tilde L_-={h_-'}^2L_- - \dfrac{1}{2} S[h_-;x^-],
	\eea
	where $S[h;x]$ is the Schwarz derivative
	\be
	S[h(x);x]=\frac{h'''}{h'}-\frac{3h''^2}{2h'^2}.
	\ee
	It is readily seen that in the infinitesimal form, where $h_\pm(x)=x^\pm+\eps_\pm(x)$, the above reduce to \eqref{delta-chi-L}.
	It is also illuminating to explicitly implement the positivity of $h'_\pm$ through
	\be
	h_\pm'= e^{\Psi_\pm},
	\ee
	where $\Psi_\pm$ are two real single-valued functions. In terms of $\Psi$ fields the Schwarz derivative takes a simple form and the expressions for $\tilde L_\pm$ become
	\be\label{Liouville-form}
	\tilde L_+[\Psi_+,L_+]= e^{2\Psi_+} L_+(x^+)+\frac{1}{4}\Psi_+'^2-\frac{1}{2}\Psi_+'',\qquad
	\tilde L_-[\Psi_-,L_-]= e^{2\Psi_-} L_-(x^-)+\frac{1}{4}\Psi_-'^2-\frac{1}{2}\Psi_-''.
	\ee
	This reminds the form  of a Liouville stress-tensor and dovetails with the fact that AdS$_3$ gravity with Brown-Henneaux boundary conditions may be viewed as a Liouville theory \cite{Coussaert:1994tu} (see also  \cite{Kim:2015qoa} for a recent discussion).

We finally note that not all functions $h_\pm$ generate new solutions. The solutions to $\tilde L_+ = L_+$, $\tilde L_- = L_-$ are coordinate transformations which leave the fields invariant: they are finite transformations whose infinitesimal versions are generated by the isometries. There are therefore some linear combinations of symplectic symmetries which do not generate any new charges. These ``missing'' symplectic charges are exactly compensated by the charges associated with the Killing vectors that we will discuss in section \ref{Killing-sec}.

\subsection{Phase space in Gaussian null coordinates}\label{retarded-coord-sec}

	In working out the symplectic symmetry generators, their charges and their algebra we used Fefferman-Graham coordinates which are very well adapted in the holographic description. One may wonder if similar results may be obtained using different coordinate systems. This question is of interest because the symplectic symmetries \eqref{symplectic symmetries} were obtained as the set of infinitesimal diffeomorphisms  which preserved the Fefferman-Graham condition \eqref{F-G gauge} and one may wonder whether the whole phase space and symplectic symmetry setup is dependent upon that particular choice.
	
	Another coordinate frame of interest may be defined a Gaussian null coordinate system,
	\bea
	g_{rr} = 0,\qquad g_{ru} = -1,\qquad g_{r\phi} = 0,
	\eea
in which $\p_r$ is an everywhere null vector field. We note along the way that the $\ell \rightarrow \infty$ limit can be made well-defined in this coordinate system after a careful choice of the scaling of other quantities \cite{Barnich:2012aw}. This leads to the BMS$_3$ group and phase space.
	
	The set of all locally AdS$_3$ geometries subject to Dirichlet boundary conditions with flat cylindrical boundary metric in such coordinate system takes the form
	\bea\label{Banados-retarted}
	ds^2 = \left( -\frac{r^2}{\ell^2}+2\ell M(u^+,u^-) \right)du^2 -2 du dr +2\ell J(u^+,u^-) du d\phi +r^2 d\phi^2,
	\eea
	where $u^\pm=u/\ell\pm \phi$. Requiring \eqref{Banados-retarted} to be solutions to AdS$_3$ Einstein's equations \eqref{action-eom} implies
	\bea\label{M-J-Lpm}
	\ell M(u^+,u^-) = L_+(u^+)+L_-(u^-), \qquad J(u^+,u^-) = L_+(u^+)-L_-(u^-).
	\eea

	As in the Fefferman-Graham coordinates, one may then view the set of metrics $g$ in \eqref{Banados-retarted}  and generic metric perturbations within the same class $\delta g$ (i.e. metrics with $L_\pm\to L_\pm+\delta L_\pm$) as members of an on-shell phase space and its tangent space. Since the coordinate change between the two Fefferman-Graham and Gaussian null coordinate systems is field dependent, the presymplectic form  cannot be directly compared between the two. After direct evaluation, we note here that both the Lee-Wald and the invariant presymplectic forms vanish on-shell
	\bea
	\boldsymbol \omega^{LW}[\delta g,\delta g ; g] \approx 0 ,\qquad \boldsymbol \omega^{inv}[\delta g,\delta g ; g] \approx 0,
	\eea
	since the boundary term which relates them vanishes off-shell, $\boldsymbol E[\delta g ,\delta g ; g] = 0$. This implies in particular that the conserved charges defined from either presymplectic form (either Iyer-Wald or Barnich-Brandt charges) will automatically agree.
	
\subsection{Symplectic symmetries and charges in Gaussian null coordinates}\label{retarded-charges-sec}
	
	The phase space of metrics in Gaussian null coordinate system \eqref{Banados-retarted} is preserved under the action of the vector field $\xi$
	\bea\label{retarded-symp-sym}
	\xi &=& \frac{1}{2} \bigg\{ \ell(Y_++Y_-)\p_u+ \big( (Y_+-Y_-)-\frac{\ell}{r}(Y_+'-Y_-') \big) \p_\phi\nn \\
	&+& \left( -r (Y_+'+Y_-')+\ell (Y_+''+Y_-'') -\frac{\ell^2}{ r} (L_+-L_-)(Y_+'-Y_-')\right)\p_r \bigg\},
	\eea
	where $Y_+=Y_+(u^+)$, $Y_- = Y_-(u^-)$. More precisely, we have
	\bea
	\delta_\xi g=\cL_\xi g_{\mu\nu}(L_+,L_-)  = g_{\mu\nu}(L_++\delta_\xi L_+ ,L_- + \delta_\xi L_- )-g_{\mu\nu}(L_+,L_-),
	\eea
	where
	\bea\label{Virl}
	\delta_\xi L_{\pm} = Y_\pm \p_\pm L_\pm +2 L_\pm Y_\pm'-\frac{1}{2} Y_\pm''',
	\eea
	stating that (Fourier modes of) $L_\pm$ are related to generators of a Virasoro algebra.
	
	It is easy to show that the surface charge one-forms are integrable $\boldsymbol k_\xi[\delta g ; g] = \delta ( \boldsymbol Q_\xi[g])$ in the phase space. The surface charge one-forms are determined up to boundary terms. It is then convenient  to subtract the following subleading boundary term at infinity,
	\bea
	\boldsymbol B_\xi =\frac{\ell^2}{32\pi G}\left( \frac{1}{r}(L_+-L_-)(Y_+ + Y_-) \right) \label{bnd1}
	\eea
	so that the total charge $\boldsymbol Q'_\xi$ is given by the radius independent expression
	\bea\label{oneV}
	\boldsymbol Q'_\xi \equiv \boldsymbol Q_\xi - \boldsymbol d \boldsymbol B_\xi = \frac{\ell}{8\pi G} (  L_+ Y_+ du^+ - L_- Y_- du^- ).
	\eea
	The two sets of Virasoro charges can then be obtained by integration on the circle spanned by $\phi$. They obey the centrally extended Virasoro algebra under the Dirac bracket as usual, as a consequence of \eqref{Virl}. Since the result is exact, the Virasoro charges and their algebra is defined everywhere in the bulk. The symplectic symmetry generators $\xi$ are field dependent (i.e. they explicitly depend on $L_\pm$), and hence their algebra is closed once we use the adjusted bracket defined in subsection \ref{modified-bracket-sec}. Also note that in the reasoning above we did not use the fact that $\phi$ is periodic until the very last step where the Virasoro charges are defined as an integral over the circle. If instead the coordinate $\phi$ is not periodic, as it is relevant to describe AdS$_3$ with a planar boundary, the Virasoro charge can be replaced by \textit{charge densities}, defined as the one-forms \eqref{oneV}.
	
	We conclude this section with the fact that both phase spaces constructed above in Fefferman-Graham coordinates and Gaussian null coordinates are spanned by two holomorphic functions and their symmetry algebra and central extension are the same. This implies that there is a one-to-one map between the two phase spaces, and therefore the corresponding holographic dynamics (induced by the Dirac bracket) is not dependent upon choosing either of these coordinate systems. We shall return to this point in the discussion section.

\section{The two Killing symmetries and their charges}\label{Killing-sec}
	
So far we discussed the symplectic symmetries of the phase space. These are associated with non-vanishing metric perturbations which are degenerate directions of the on-shell presymplectic form. A second important class of symmetries are the Killing vectors which are associated with vanishing metric perturbations. In this section we analyze these vector fields, their charges and their commutation relations with the symplectic symmetries. We will restrict our analysis to the case of asymptotically globally AdS$_3$ where $\phi$ is $2\pi$-periodic. We use Fefferman-Graham coordinates for definiteness but since Killing vectors are geometrical invariants, nothing will depend upon this specific choice.

\subsection{Global Killing vectors}\label{Global-Killings-sec}
	
Killing vectors are vector fields along which the metric does not change. All diffeomorphisms preserving the Fefferman-Graham coordinate system are generated by the vector fields given in \eqref{symplectic symmetries}. Therefore, Killing vectors have the same form as $\chi$'s, but with the extra requirement that $\delta L_\pm$ given by \eqref{delta-chi-L} should vanish. Let us denote the functions $\eps_\pm$ with this property by $K_\pm$ and the corresponding Killing vector by $\zeta$ (instead of $\chi$). Then, $\zeta$ is a Killing vector if and only if
	\begin{align}\label{stabilizer}
		K_+'''-4L_+ K_+'-2K_+ L_+' =0,\qquad K_-'''-4L_- K_-'-2K_- L_-' =0.
	\end{align}
	These equations were thoroughly analyzed in \cite{Sheikh-Jabbari:2014nya}  and  we only provide a summary of the results relevant for our study here.
	The above linear third order differential equations have three linearly independent solutions and hence Ba\~nados geometries in general have six (local) Killing vectors which form an $sl(2,\mathbb{R})\times sl(2,\mathbb{R})$ algebra, as  expected. The three solutions take the form $K_+ = \psi_i \psi_j$, $i,j=1,2$ where $\psi_{1,2}$ are the two independent  solutions to the second order Hill's equations
	\bea
	\psi'' = L_+(x^+) \psi
	\eea
	where $L_+(x^++2\pi)=L(x^+)$. Therefore, the function $K_+$ functionally depends upon $L_+$  but not on $L_+'$, i.e. $K_+=K_+(L_+)$. This last point will be crucial for computing the commutation relations and checking integrability as we will shortly see. The same holds for the right moving sector. In general, $\psi_i$ are not periodic functions under $\phi \sim \phi+2\pi$ and therefore not all six vectors described above are global Killing vectors of the geometry. However, Floquet's theorem \cite{magnus2013hill} implies that the combination $\psi_1 \psi_2$ is necessarily periodic. This implies that Ba\~nados geometries have at least two global Killing vectors.
	Let us denote these two global Killing vectors by $\zeta_\pm$,
	\bea
	\zeta_+=\chi(K_+, K_-=0;L_\pm),\qquad  \zeta_-=\chi(K_+=0, K_-; L_\pm),
	\eea
	where $\chi$ is the vector field given in \eqref{symplectic symmetries}. These two vectors define two global $U(1)$ isometries of Ba\~nados geometries.

The important fact about these global $U(1)$ isometry generators is that they commute with \emph{each} symplectic symmetry generator $\chi$ \eqref{symplectic symmetries}: Since the vectors are field-dependent, one should use  the adjusted bracket \eqref{modified-bracket} which reads explicitly as
\begin{align}
		\big[\chi(\eps;L),\zeta(K;L)\big]_*&=\big[\chi(\eps;L),\zeta(K;L)\big]_{L.B.}-\Big(\de^L_\eps \zeta(K;L)-\de^L_{K} \chi(\eps;L)\Big),\nn
\end{align}
where the first term on the right-hand side is the usual Lie bracket. Since $K=K(L)$, the adjustment term reads as
	\begin{align}
		\de^L_\eps \zeta(K(L);L)&=\de_\eps L\;\dfrac{\p}{\p L}\zeta(K;L) +\zeta(\de^L_\eps K;L), \label{eq:99}\\
		\de^L_{K} \chi(\eps;L) &= \de_K L\;\dfrac{\p}{\p L} \chi(\eps;  L) = 0
	\end{align}
	where we used the fact that $\zeta,\chi$ are linear in their first argument as one can see from \eqref{symplectic symmetries} and we used Killing's equation. We observe that we will get only one additional term with respect to the previous computation \eqref{bracket} due to the last term in \eqref{eq:99}. Therefore,
	\begin{align}
		\big[\chi(\eps;L),\zeta(K(L);L)\big]_*&= \zeta(\eps\, K' - \eps' \,K;L) -\zeta(\de^L_\eps K;L).
	\end{align}
	Now the variation of Killing's condition \eqref{stabilizer} implies that
	$$
	(\de K)'''-4L(\de K)'-2L'\de K=4\delta L K'+2(\delta L)'K.
	$$
	Then, recalling \eqref{delta-chi-L} and using again \eqref{stabilizer} we arrive at
	\begin{align}
		\de^L_\eps K&=\eps\, K' - \eps' \,K\label{deltaK},
	\end{align}
	and therefore
	\be\label{chi-zeta-commute}
	\big[\chi(\eps;L),\zeta(K(L);L)\big]_*=0.
	\ee
	The above may be intuitively understood as follows. $\zeta$ being a Killing vector field does not transform $L$, while a generic $\chi$ transforms $L$. Now the function $K$ is a specific function of the metric, $K=K(L)$. The adjusted bracket is defined such that it removes the change in the metric and only keeps the part which comes from Lie bracket of the corresponding vectors as if $L$ did not change.

It is interesting to compare the global Killing symmetries and the symplectic symmetries. The symplectic symmetries are given by \eqref{symplectic symmetries} and determimned by functions $\eps_\pm$. The functions $\eps_\pm$ are field independent, so that they are not transformed by moving in the phase space. On the other hand, although the Killing vectors have the same form \eqref{symplectic symmetries}, their corresponding functions $\eps_\pm$ which are now denoted by $K_\pm$, are field dependent as a result of \eqref{stabilizer}. Therefore the Killing vectors differ from one geometry to another. Accordingly if we want to write the Killing vectors in terms of the symplectic symmetry Virasoro modes $\chi_n^\pm$ \eqref{Witt}, we have
		\begin{align}
			\zeta_+=\sum_n c^+_n(L_+) \chi_n^+,\qquad \qquad 	\zeta_-=\sum_n c^-_n(L_-) \chi_n^-.
					\end{align}	
For example for a BTZ black hole, one can show using \eqref{stabilizer} that the global Killing vectors are $\zeta_\pm=\chi_0^\pm$ while for a BTZ black hole with Virasoro hair or ``BTZ Virasoro descendant'', which is generated by the coordinate transformations in section \ref{finite-transf-sec}, it is a complicated combination of Virasoro modes. For the case of global AdS$_3$ with $L_\pm=-\frac{1}{4}$ (but not for its descendents), \eqref{stabilizer} implies that there are six global Killing vectors which coincide with the subalgebras $\{\chi^+_{1,0,-1}\}$ and $\{\chi^-_{1,0,-1}\}$ of symplectic symmetries.

We close this part by noting the fact that although we focused on single-valued $K$ functions, one may readily check that this analysis and in particular \eqref{chi-zeta-commute} is true for any $K$ which solves \eqref{stabilizer}. Therefore, all six generators of local $sl(2,\mathbb{R})\times sl(2,\mathbb{R})$ isometries commute with symplectic symmetry generators $\chi$ \eqref{symplectic symmetries}. This is of course expected as all geometries \eqref{Banados} are locally $sl(2,\mathbb{R})\times sl(2,\mathbb{R})$ invariant. {We shall discuss this point further in section \ref{discussion-sec}.}

\subsection{Conserved charges associated with the $U(1)$ Killing vectors}\label{Jpm-sec}

Similarly to the Virasoro charges \eqref{chk}, the infinitesimal charges associated to Killing vectors can be computed using \eqref{kgrav}, leading to
	\be\label{delta-J}
	\delta J_+  =\frac{\ell}{8\pi G} \int_0^{2\pi}  d\phi\ K_+(L_+) \delta L_+,\qquad \delta J_-  =\frac{\ell}{8\pi G} \int_0^{2\pi}  d\phi\ K_-(L_-) \delta L_-.
	\ee
	\paragraph{Integrability of Killing charges.}
Given the field dependence of the $K$-functions, one may inquire about the integrability of the charges $J_\pm$ over the phase space. In appendix \ref{integrability-appendix-sec}, we find the necessary and sufficient condition for the integrability of charges associated with field dependent vectors. However, in the present case, the integrability of $J_\pm$ can be directly checked as follows
	\begin{align}\label{ci}
		\delta_1 (\delta_2 J) &= \frac{\ell}{8\pi G}\oint \delta_1 K(L) \; \delta_2 L=\frac{\ell}{8\pi G} \oint \dfrac{\p K}{\p L}\; \delta_1 L\; \delta_2 L,
	\end{align}
	and therefore  $\delta_1 (\delta_2 J )- \delta_2 (\delta_1 J )=0$.
	
Having checked the integrability, we can now proceed with finding the explicit form of charges through an integral along a suitable path over the phase space connecting a reference field configuration to the configuration of interest. However, as we will see in section \ref{Banados-Orbits-sec}, the Ba\~nados phase space is not simply connected and therefore one cannot reach any field configuration through a path from a reference field configuration. As a result, the charges should be defined independently over each connected patch of the phase space. In section \ref{Banados-Orbits-sec} we will give the explicit form of charges over a patch of great interest, i.e. the one containing BTZ black hole and its descendants. We then find a first law relating the variation of entropy to the variation of these charges.
	
	\paragraph{Algebra of Killing and symplectic charges.} We have already shown in section \ref{Global-Killings-sec} that the adjusted bracket between generators of respectively symplectic and Killing symmetries vanish. If the charges are correctly represented, it should automatically follow that the corresponding charges $L_n, J_+$ (and $\bar L_n, J_-$) also commute:
	\bea\label{J-L-commute}
	\{ J_\pm , L_n \} = \{ J_\pm , \bar L_n \} = 0.
	\eea
	Let us check \eqref{J-L-commute}. By definition we have
	\bea
	\{ J_+ , L_n \}  = -\delta_K L_n,
	\eea
	where one varies the dynamical fields in the definition of $L_n$ with respect to the Killing vector $K$. Since $K$ leaves the metric unchanged, we have $\delta_K L_+(x^+) = 0$ and therefore directly $\delta_K L_n = 0$. Now, let us also check that the bracket is anti-symmetric by also showing
	\bea\label{J-fixed-on-orbit}
	\{ L_n , J_+ \} \equiv  - \delta_{L_n} J_+ = 0.
	\eea
	This is easily shown as follows:
	\bea\label{Jvar}
	\delta_{L_n} J_+ &=& \frac{\ell}{8\pi G} \int_0^{2\pi} d\phi\ K_+ \delta_{\eps^+_n} L
	= \frac{\ell}{8\pi G} \int_0^{2\pi} d\phi\ K_+ (\eps^+_n L_+' +2 L_+ \eps^+_n{'}-\frac{1}{2}\eps^+_n{'''} )\nnr
	&=&  \frac{\ell}{8\pi G} \int_0^{2\pi} d\phi\ (-L_+' K_+ -2L_+ K_+' +\frac{1}{2}K_+''')\eps_{+,n}
	= 0
	\eea
	after using \eqref{delta-chi-L}, integrating by parts and then using \eqref{stabilizer}. The same reasoning holds for $J_-$ and $\bar L_n$.

In general, the Ba\~nados phase space only admits two Killing vectors. {An exception} is the descendants of the vacuum AdS$_3$ which admit six globally defined Killing vectors. In that case, the two $U(1)$ Killing charges are $J_\pm = -\frac{1}{4}$ and the other four $\frac{SL(2,\mathbb R)}{U(1)} \times \frac{SL(2,\mathbb R)}{U(1)}$ charges are identically zero. In the case of the decoupled near-horizon extremal phase space defined in section \ref{extremal-NH-sec} we will have four global Killing vectors with the left-moving $U(1)_+$ charge $J_+$ arbitrary, but the $SL(2,\mathbb R)_-$ charges all vanishing $J^a_-=0$, $a=+1,0,-1$.

\section{Phase space as Virasoro coadjoint orbits}\label{Banados-Orbits-sec}

As discussed in the previous sections, one can label each element of the phase space in either Fefferman-Graham coordinates  or Gaussian null coordinates, described respectively by \eqref{Banados} and \eqref{Banados-retarted}, by its symplectic charges $L_n, \bar L_n$ and its global commuting Killing charges $J_\pm$. Moreover, the phase space functions $L_\pm$ transform under the coadjoint action of the Virasoro algebra, see \eqref{delta-chi-L}. Hence, we are led to the conclusion that the phase space forms a reducible representation of the Virasoro group composed of distinct Virasoro coadjoint orbits.

Construction of Virasoro coadjoint orbits has a long and well-established literature, see e.g. \cite{Balog:1997zz} and references therein. In this literature the $\delta L_\pm=0$ (i.e. \eqref{stabilizer}) equation is called the stabilizer equation \cite{Witten:1987ty} and specifies the set of transformations which keeps one in the same orbit. The stabilizer equation and classification of its solutions is hence the key to the classification of Virasoro coadjoint orbits. Since an orbit is representation of the Virasoro group it might as well be called a conformal multiplet. The elements in the same orbit/conformal multiplet may be mapped to each other upon the action of coordinate transformations \eqref{finite-L-transf}. Explicitly, a generic element/geometry in the same orbit (specified by $\tilde L_\pm$) is related to a single element/geometry with  $L_\pm$ given as \eqref{Liouville-form} for arbitrary periodic functions $\Psi_\pm$.  One canhence  classify the orbits by the set of periodic functions $L_\pm(x^\pm)$ which may not be mapped to each other through \eqref{Liouville-form}. One may also find a specific $L_\pm$, the \emph{representative of the orbit}, from which one can generate the entire orbit by conformal transformations \eqref{Liouville-form}. In the language of a dual 2d CFT, each orbit may be viewed as a primary operator together with its conformal descendants. Each geometry is associated with (one or many) primary operators or descendants thereof, in the dual 2d CFT. From  this discussion it also follows that there is no regular coordinate transformation respecting the chosen boundary conditions, which  moves us among the orbits.

\subsection{Classification of Virasoro orbits}\label{Virasoro-orbits-sec}

Let us quickly summarize some key results from \cite{Balog:1997zz}. In order to avoid notation clutter we focus on a single sector, say the $+$ sector (which we refer to as left-movers).
One may in general distinguish two classes of orbits: those where a constant representative exists and those where it doesn't. The constant $L_+$ representatives correspond to the better studied geometries, e.g. see \cite{Garbarz:2014kaa,Sheikh-Jabbari:2014nya}  for a review. They fall into four categories:
\begin{itemize}
\item
\emph{Exceptional orbits} ${\cal E}_n$ with representative $L=-n^2/4,\  n=0,1, 2,3,\cdots$. The orbit ${\cal E}_0 \times {\cal E}_0$ admits the zero mass BTZ as a representative. (The $n=0$ case coincides with the hyperbolic orbit ${\cal B}_0(0)$, see below.) The ${\cal E}_1 \times {\cal E}_1$ orbit admits global AdS$_3$ as a representative and therefore corresponds to the vacuum Verma module in the language of a 2d CFT on the cylinder. For $n \geq 2$, ${\cal E}_n \times {\cal E}_n$ is represented by an $n$-fold cover of global AdS$_3$.
\item \emph{Elliptic orbits} ${\cal C}_{n, \nu}$, with representative $L=-(n+\nu)^2/4$, $n=0,1,\dots$ and $0<\nu<1 $.  The geometries with both $L_\pm$ elliptic orbit representatives with $n=0$ correspond to conic spaces, particles on AdS$_3$ \cite{Deser:1983nh} and geometries in this orbit may be viewed as ``excitations'' (descendants) of particles on the AdS$_3$.
 \item \emph{Hyperbolic orbits} ${\cal B}_0 (b)$, with representative $L=b^2/4$, where $b$ is generic real non-negative number $b\geq 0$. The $b=0$ case coincides with the ${\cal E}_0$ orbit. The geometries with both $L_\pm=b_\pm^2/4$ are BTZ black holes. The special case of $b_\pm=0$ is the massless BTZ and $b_-=0$ is extremal BTZ.
\item \emph{Parabolic orbit} ${\cal P}_0^+$, with representative $L=0$. The geometries associated with ${\cal P}_0^+\times {\cal B}_0 (b)$ orbits correspond to the self-dual orbifold \cite{Coussaert:1994tu} which may also be obtained as the near horizon geometry of extremal BTZ black holes. In particular, ${\cal P}_0^+\times {\cal B}_0 (0)$ corresponds to null selfdual orbifold \cite{Balasubramanian:2003kq}. The
${\cal P}_0^+\times{\cal P}_0^+$ orbit corresponds to AdS$_3$ in the Poincar\'e patch and its descendants, which in the dual 2d CFT corresponds to vacuum Verma module of the CFT on 2d plane.

\end{itemize}

The non-constant representative orbits, come into three categories, the generic \emph{hyperbolic} orbits ${\cal B}_n(b)$ and two \emph{parabolic} orbits ${\cal P}^\pm_n$, $n\in \mathbb{N}$. Geometries associated with these orbits are less clear and understood. This question will be addressed in a future publication \cite{Sheikh-Jabbari:2015nn}.

To summarize, if we only focus on the labels on the orbits, the ${\cal E}_n, {\cal P}^\pm_n$ orbits have only an integer label, the ${\cal C}_{n,\nu}$ is labeled by a real number between 0 and 1 and an integer, and the hyperbolic ones ${\cal B}_n(b)$ with an integer and a real positive number.

\subsection{ Killing charges as orbit labels }\label{subsec: Killing labels}

As shown in \eqref{Jvar}, all the geometries associated with the same orbit have the same  $J_\pm$ charges. In other words, $J_\pm$ do not vary as we make coordinate transformations using $\chi$ diffeomorphisms \eqref{symplectic symmetries}; $J_\pm$ are ``orbit invariant'' quantities. One may hence relate them with the labels on the orbits, explicitly, $J_+$ should be a function of $b$ or $\nu$ for the hyperbolic or elliptic orbits associated to the left-moving copy of the Virasoro group and $J_-$ a similar function of labels on the right-moving copy of the Virasoro group.

The Ba\~nados phase space has a rich topological structure. It consists of different disjoint patches. Some patches (labeled by only integers) consist of only one orbit, while some consist of a set of orbits with a continuous parameter. On the other hand, note that the conserved charges in covariant phase space methods are defined through an integration of infinitesimal charges along a path connecting a reference point of phase space to a point of interest. Therefore, the charges can be defined only over the piece of phase space simply connected to the reference configuration. For other patches, one should use other reference points. In this work we just present explicit analysis for the  ${\cal B}_0(b_+)\times  {\cal B}_0(b_-)$ sector of the phase space. Since this sector corresponds to the family of BTZ black holes of various mass and angular momentum and their descendants, we call it the BTZ sector. Note that there is no regular coordinate transformation respecting the chosen boundary conditions, which  moves us among the orbits. In particular for the BTZ sector, this means that there is no regular coordinate transformation which relates BTZ black hole geometries with different mass and angular momentum, i.e. geometries with different $b_\pm$.

We now proceed with computing the charges $J_\pm$ for an arbitrary field configuration in the BTZ sector of the phase space. Since the charges are integrable, one can choose any path from a reference configuration to the desired point. We fix the reference configuration to be the massless BTZ with $L_\pm = 0$. We choose the path to pass by the constant representative $L_\pm$ of the desired solution of interest $\tilde L_\pm(x^\pm)$. Let us discuss $J_+$ (the other sector follows the same logic). Then the charge is defined as
\begin{align}
J_+ & =\int_\gamma \delta J_+ =\int_{0}^{\tilde L_+} \delta J_+  = \int_{0}^{L_+} \delta J_+ + \int_{L_+}^{\tilde L_+} \delta J_+.
\end{align}
We decomposed the integral into two parts: first the path \emph{across} the orbits, between constant representatives $L_+ = 0$ and $L_+$ and second the path along (within) a given orbit with representative $L_+$. Since the path along the orbit does not change the values $J_\pm$ ($\delta_\chi J_\pm=0$), the second integral is zero.  Accordingly, the charge is simply given by
\begin{align}
J_+&= \frac{\ell}{8\pi G}\int_{0}^{L_+} \oint d\varphi K_+(L)\de L
\end{align}
where $L_+$ is a constant over the spacetime. Solving \eqref{stabilizer} for constant $L_\pm$ and assuming periodicity of $\phi$, we find that $K_\pm=const$. Therefore the Killing vectors are $\p_\pm$ up to a normalization constant, which we choose to be 1. Hence $K_+(L)=1$, and
\begin{align}\label{defJp}
J_+&=\frac{\ell}{4 G} L_+ ,\qquad J_-=\frac{\ell}{4 G} L_-.
\end{align}
Therefore the Killing charges are a multiple of the Virasoro zero mode of the constant representative.

\subsection{Thermodynamics of Ba\~nados geometries}\label{Thermodynamics-sec}

Since the BTZ descendants are obtained through a finite coordinate transformation from the BTZ black hole, the descendants inherit the causal structure and other geometrical properties of the BTZ black hole. We did not prove that the finite coordinate transformation is non-singular away from the black hole Killing horizon but the fact that the Virasoro charges are defined all the way to the horizon gives us confidence that there is no singularity between the horizon and the spatial boundary. The geometry of the Killing horizon was discussed in more detail in \cite{Sheikh-Jabbari:2014nya}.

The area of the outer horizon defines a geometrical quantity which is invariant under diffeomorphisms. Therefore the BTZ descendants admit the same area along the entire orbit. The angular velocity and surface gravity are defined geometrically as well, given a choice of normalization at infinity. This choice is provided for example by the asymptotic Fefferman-Graham coordinate system which is shared by all BTZ descendants. Therefore these chemical potentials $\tau_\pm$ are also orbit invariant and are identical for all descendants and in particular are constant. This is the zeroth law for the BTZ descendant geometries.

One may define more precisely $\tau_\pm$ as the chemical potentials conjugate to $J_\pm$ \cite{Sheikh-Jabbari:2015nn}. Upon varying the parameters of the solutions we obtain a linearized solution which obey the first law
\be\label{first-law}
\delta S= \tau_+ \delta J_++\tau_- \delta J_- .
\ee
This first law is an immediate consequence of the first law for the BTZ black hole since all quantities are geometrical invariants and therefore independent of the orbit representative. In terms of $L_\pm$, the constant representatives of the orbits in the BTZ sector, one has {\eqref{defJp} and \cite{Kraus:2006wn} }
\be
\tau_\pm=\frac{\pi}{\sqrt{L_\pm}}
\ee
and the entropy takes the usual Cardy form
\bea
S=\frac{\pi}{3} c(\sqrt{L_+}+\sqrt{L_-}) .
\eea
One can also write the Smarr formula in terms of orbit invariants as
\bea
S = 2(\tau_+ J_++\tau_- J_-).
\eea

The only orbits which have a continuous label (necessary to write infinitesimal variations) and which admit a bifurcate Killing horizon are the hyperbolic orbits \cite{Sheikh-Jabbari:2014nya, Sheikh-Jabbari:2015nn}.
The extension of the present discussion to generic hyperbolic orbits (and not just for the BTZ sector) will be postponed to \cite{Sheikh-Jabbari:2015nn}.

\section{Extremal phase space and decoupling limit}\label{extremal-NH-sec}

We define the ``extremal phase space'' as the subspace of the set of all Ba\~nados geometries (equipped with the invariant presymplectic form) with the restriction that the right-moving function $L_-$ vanishes identically. The Killing charge $J_-$ is therefore identically zero. Also, perturbations tangent to the extremal phase space obey $\delta L_- =0$ but $\delta L_+$ is an arbitrary left-moving function.

A particular element in the extremal phase space is the extremal BTZ geometry with $M \ell=J$. It is well-known that this geometry admits a decoupled near-horizon limit which is given by the self-dual spacelike orbifold of AdS$_3$ \cite{Coussaert:1994tu}
\bea\label{sdo}
ds^2 = \frac{\ell^2}{4} \left(-r^2 dt^2 + \frac{dr^2}{r^2} + \frac{4 |J|}{k} (d\phi - \frac{r}{{2}\sqrt{|J|/k}}dt )^2 \right),\qquad \phi \sim \phi + 2\pi,
\eea
where $k \equiv \frac{\ell}{4G}$. A Virasoro algebra exists as asymptotic symmetry in the near-horizon limit and this Virasoro algebra has been argued to be related to the asymptotic Virasoro algebra defined close to the AdS$_3$ spatial boundary \cite{Balasubramanian:2009bg}. Since these asymptotic symmetries are defined at distinct locations using boundary conditions it is not entirely obvious that they are uniquely related. Now, using the concept of symplectic symmetries which extend the asymptotic symmetries to the bulk spacetime, one deduces that the extremal black holes are equipped with one copy of Virasoro hair. The Virasoro hair transforms under the action of the Virasoro symplectic symmetries, which are also defined everywhere outside of the black hole horizon.

One subtlety is that the near-horizon limit is a decoupling limit obtained after changing coordinates to near-horizon comoving coordinates. We find two interesting ways to take the near-horizon limit. In Fefferman-Graham coordinates the horizon is sitting at $r=0$ and it has a constant angular velocity $1/\ell$ \emph{independently of the Virasoro hair}. Therefore taking a near-horizon limit is straightforward and one readily obtains the near-horizon Virasoro symmetry. It is amusing that the resulting vector field which generates the symmetry differs from the ansatz in \cite{Balasubramanian:2009bg}, as well as the original Kerr/CFT ansatz \cite{Guica:2008mu} and the newer ansatz for generic extremal black holes \cite{Compere:2014cna, Compere:2015mza}. The difference is however a vector field which is pure gauge, i.e. charges associated with it are zero.

A second interesting way to take the near-horizon limit consists in working with coordinates such that the horizon location depends upon the Virasoro hair. This happens in Gaussian null coordinates. Taking the near-horizon limit then requires more care. This leads to a yet different Virasoro ansatz for the vector field which is field dependent. After working out the details, a chiral half of the Virasoro algebra is again obtained, which also shows the equivalence with the previous limiting procedure.

\subsection{Decoupling limit in Fefferman-Graham coordinates}

The general metric of the extremal phase space of AdS$_3$ Einstein gravity with Brown-Henneaux boundary conditions and in the Fefferman-Graham coordinate system is given by
\begin{align}\label{extremal Banados}
ds^2=\dfrac{\ell^2}{r^2}dr^2-r^2dx^+dx^- + \ell^2L(x^+){dx^+}^2,\qquad x^\pm = t/\ell \pm \phi,\qquad \phi \sim \phi+2\pi
\end{align}
{where we dropped the $+$ subscript, $L_+=L$}. It admits two global Killing vectors: $\p_-$ and $\zeta_+$ defined in subsection \ref{Global-Killings-sec}. In the case of the  extremal BTZ orbit, the metrics \eqref{extremal Banados} admit a Killing horizon at $r=0$ which is generated by the Killing vector $\p_-$ \cite{Sheikh-Jabbari:2014nya}.

One may readily see that a diffeomorphism $\chi(\epsilon_+,\epsilon_-=0)$ defined from \eqref{symplectic symmetries} with  arbitrary $\epsilon_+(x^+)$, namely
\begin{align}\label{chi-ext}
\chi_{ext}=\dfrac{\ell^2\epsilon_+''}{2r^2} \p_-  + \epsilon_+ \p_+ - \half r \epsilon_+' \p_r,
\end{align}
is tangent to the phase space. Indeed, it preserves the form of the metric \eqref{extremal Banados}. Remarkably, the field dependence, i.e. the dependence on $L_+$, completely drops out in $\chi_{ext}$. Note however that although
$\chi_{ext}$ is field independent, the Killing vector $\zeta_+$ is still field dependent.
From the discussions of section \ref{Banados-gauge-sec} it immediately follows that $\chi_{ext}$ generates symplectic symmetries.

One may then take the decoupling limit
\bea
t \rightarrow \frac{\ell\,\tilde t}{\lambda},\qquad \phi \rightarrow \phi + \Omega_{ext} \frac{\ell\,\tilde t}{\lambda},{\qquad r^2\rightarrow 2\ell^2\, \lambda \tilde r},\qquad \lambda \rightarrow 0
\eea
where $ \Omega_{ext}=-1/\ell$ is the constant angular velocity at extremality. As a result $x^+ \rightarrow \phi$ and {$x^- \rightarrow 2 \frac{\tilde t}{\lambda} - \phi $}. Functions periodic in $x^+$ are hence well-defined in the decoupling limit while functions periodic in $x^-$ are not. Therefore, the full Ba\~nados phase space does not admit a decoupling limit. Only the extremal part of the Ba\~nados phase space does. Also, since $\frac{\tilde t}{\lambda}$ is dominant with respect to $\phi$ in the near-horizon limit, the coordinate $x^-$ effectively decompactifies in the limit while $x^+$ remains periodic. Since $-dx^+ dx^-$ is the metric of the dual CFT, this leads to the interpretation of the decoupling limit as a discrete-light cone quantization of the dual CFT \cite{Balasubramanian:2009bg}.

In this limit the metric \eqref{extremal Banados} and symplectic symmetry generators \eqref{chi-ext} become
\begin{align}
\dfrac{ds^2}{\ell^2}&=\dfrac{d\tilde{r}^2}{4\tilde{r}^2}-{4\tilde{r}} d\tilde td\phi+L(\phi){d\phi}^2\label{NH-metric-Banados}\\
\chi_{ext} &=\dfrac{\epsilon''(\phi)}{8\tilde{r}} \p_{\tilde t} -  \tilde{r} \epsilon'(\phi) \p_{\tilde{r}}  + \epsilon(\phi) \p_\phi \label{NH-chi-Banados},
\end{align}
{where we dropped again the $+$ subscript, $\eps_+=\eps$.} As it is standard in such limits, this geometry acquires an enhanced global $SL(2,\mathbb{R})_- \times U(1)_+$ isometry \cite{Sheikh-Jabbari:2014nya, Li:2013pra}. The $sl(2,\mathbb{R})_-$ Killing vectors are given as
\begin{align}
\xi_1=\dfrac{1}{2}\p_{\tilde t},\;\;\;\;\xi_2=\tilde t \p_{\tilde t} - \tilde{r} \p_{\tilde{r}},\;\;\;\;\xi_3=[({2\tilde t^2} + \dfrac{L}{8\tilde{r}^2})\p_{\tilde t} + \dfrac{1}{2\tilde{r}}\p_\phi - {4\tilde  t \tilde{r}} \p_{\tilde{r}}].
\end{align}
and obey the algebra
\begin{align}\label{sl2R-algebra}
[\xi_1,\xi_2]=\xi_1,\qquad
[\xi_1,\xi_3]=2\xi_2,\qquad
[\xi_2,\xi_3]=\xi_3,
\end{align}
The $u(1)_+$ is still generated by $\zeta_+$.

As it is explicitly seen from the metric \eqref{NH-metric-Banados}, absence of Closed Timelike Curves (CTC) requires $L(\phi) \geq 0$. This restricts the possibilities for orbits which admit a regular decoupling limit. The obvious example is the extremal BTZ orbit for which the decoupling limit is a near-horizon limit. Representatives of these orbits are the extremal BTZ black holes with $L_+ \geq 0$ constant and the near-horizon metric  \eqref{NH-metric-Banados} is precisely the self-dual orbifold \eqref{sdo} after recognizing $J = \frac{\ell}{4G}L = \frac{c}{6}L$ and setting $\tilde t=\sqrt{L_+} t/4$ and $\tilde r=r$.\footnote{For the case of the massless BTZ, one should note that there are two distinct near-horizon limits; the first leads to null self-dual orbifold of AdS$_3$ and the second to the pinching AdS$_3$ orbifold
\cite{deBoer:2010ac}.}

From the analysis provided in \cite{Balog:1997zz} one can gather that all orbits other than the hyperbolic ${\cal B}_0(b)$  and the parabolic ${\cal P}^+_0$ orbits, admit a function $L(\phi)$ which can take negative values. The corresponding geometries therefore contain CTCs. The only regular decoupling limit is therefore the near-horizon limit of generic extremal BTZ (including massless BTZ \cite{deBoer:2010ac}).  Therefore, the near-horizon extremal phase space is precisely the three-dimensional analogue of the phase space of more generic near-horizon extremal geometries discussed in \cite{Compere:2015bca, Compere:2015mza}. In other words, geometries of the form \eqref{NH-metric-Banados} which are free of CTCs are in  ${\cal P}^+_0\times  {\cal P}^+_0$ or  ${\cal P}^+_0\times {\cal B}_0(b),\ b\geq 0 $ orbits.

Under the action of $\chi_{ext}$ above, one has
\begin{align}
\delta_\chi L(\phi)= \epsilon L(\phi)' + 2 L(\phi) \epsilon' - \half \epsilon'''
\end{align}
in the decoupling limit. With the mode expansion $\epsilon=e^{i n \phi}$, one may define the symplectic symmetry generators $l_n$ which satisfy the Witt algebra,
\begin{align}
i[l_m,l_n]=(m-n)l_{m+n}.
\end{align}
The surface charge is integrable and given by
\be
H_\chi[\Phi]=\dfrac{\ell}{8 \pi G}\oint d\phi \, \epsilon(\phi) L(\phi).
\ee
Moreover, one may show that the surface charges associated to the $SL(2,\mathbb{R})_-$ Killing vectors, $J_-^a$, vanish. Interestingly, we find that the $\tilde t$ and $\tilde{r}$ components of $\chi_{ext}$ \eqref{NH-chi-Banados} do not contribute to the surface charges. The various ansatzes described in \cite{Balasubramanian:2009bg,Guica:2008mu,Compere:2014cna, Compere:2015mza} which differ precisely by the $\p_{\tilde t}$ term are therefore physically equivalent to the one in  \eqref{NH-chi-Banados}.

One may also work out the algebra of charges $H_n$ associated with $\epsilon=e^{i n \phi}$:
\be
\{H_m,H_n\}=(m-n)H_{m+n}+\dfrac{c}{12}m^3\delta_{m+n,0},
\ee
where $c$ is the usual Brown-Henneaux central charge.

The charge $J_+$ associated with the Killing vector $\zeta_+$ commutes with the $H_n$'s, as discussed in general in section \ref{Jpm-sec}. Following the analysis of section \ref{Thermodynamics-sec}, one may associate an entropy $S$ and chemical potential $\tau_+$ which satisfy the first law and Smarr relation
\be\label{first-law-Smarr-extremal}
\delta S=\tau_+ \delta J_+\,,\qquad S=2\tau_+ J_+.
\ee
These are the familiar laws of ``near horizon extremal geometry (thermo)dynamics'' {presented} in \cite{Hajian:2013lna,Hajian:2014twa}.
\subsection{Near horizon limit in Gaussian null coordinates}

Let us now consider the analogue extremal phase space but in Gaussian null coordinates. It is defined from the complete phase space discussed in section \ref{retarded-coord-sec} by setting the right-moving function $L_- = 0$. The metric is
\bea\label{Extremal-retarded-metric}
ds^2 = (-\frac{r^2}{\ell^2} + 2L_+(u^+))du^2 - 2du dr +2\ell{L_+(u^+)} dud\phi+r^2 d\phi^2,
\eea
where $u^\pm=u/\ell \pm \phi, \, \phi \sim \phi+2\pi$. It depends upon a single function $L_+(u^+)$. One may analyze the isometries of metrics \eqref{Extremal-retarded-metric}. The Killing vectors are within the family of $\xi$'s \eqref{retarded-symp-sym} with $\delta_\xi L_\pm=0$ (\emph{cf.} \eqref{Virl}). Since $L_-=0$ in this family, there are three local Killing vectors associated with solutions of $Y_-'''=0$, i.e. $Y_-=1,u^-,(u^-)^2$. The first Killing vector is $\xi_1=\p_- = \frac12(\ell\p_u-\p_\phi)$. The other two   are not globally single-valued but we will display them for future use,
\be\label{sl2R-gen-GNC}
\xi_2=u^-\p_- + \frac{\ell}{2r}\p_\phi-\frac12(r-\frac{\ell^2L_+}{r})\p_r,\ \  \xi_3=(u^-)^2\p_- + u^- \frac{\ell}{r}\p_\phi + \big[\ell- u^- (r-\frac{\ell^2L_+}{r})\big]\p_r.
\ee
Together they form an $sl(2,\mathbb{R})$ algebra \eqref{sl2R-algebra}. There is also a global $U(1)_+$ associated with the $Y_+$ functions, which is  the periodic solution to $\delta_\xi L_+=0$.

The set of geometries \eqref{Extremal-retarded-metric} together with $\xi(Y_+, Y_-=0)$ (\emph{cf}. \eqref{retarded-symp-sym}) form a phase space, elements of which fall into the Virasoro coadjoint orbits. Orbits are labeled by $J_+$. We consider for simplicity only the extremal BTZ orbit. The above geometries then have a Killing horizon at variable radius $r=r_H(u^+)$, unlike the Fefferman-Graham coordinate system studied in the previous section. The function $r_H(u^+)$ is defined from the function $L_+(u^+)$ through
\be
\label{horizon}
\ell \, \frac{dr_H}{du^+}+r_H^2=\ell^2L_+(u^+)\,.
\ee
This Killing horizon is generated by the Killing vector $\p_-$. Requiring the function $r_H$ to be real imposes a constraint on the Virasoro zero mode $\int_0^{2\pi} du^+ L_+(u^+) \geq 0$ which is obeyed in the case of the hyperbolic ${\cal B}_0(b)$ orbit. It is notable that upon replacing $r_H=\ell \frac{\psi'}{\psi}$, \eqref{horizon} exactly reduces to Hill's equation $\psi'' =L_+\psi$.

Let us now perform the following near-horizon limit,
\bea
r  = r_H(u^+) +\eps \hat r,\qquad u = \frac{\hat u}{\eps}, \qquad \phi = \hat \phi +\Omega_{ext} \frac{\hat u}{\eps},\ \qquad \eps\to 0 \label{near4}
\eea
where $\Omega_{ext}= -\frac{1}{\ell}$ is the extremal angular velocity. In this limit $u^+=\hat\phi$ is kept finite. The metric takes the form
\bea
ds^2 = -2 d\hat r d\hat u  - 4\hat r \frac{r_H(\hat\phi)}{\ell}d\hat u d\hat \phi + r_H^2(\hat\phi) d\hat \phi^2 .\label{NBTZ4}
\eea
Note also that $r_H(\hat{\phi})$ is a function of $L_+(\hat{\phi})$, as is given in \eqref{horizon}. For constant $r_H$ the metric \eqref{NBTZ4} is the self-dual AdS$_3$ metric. In general, it admits a $SL(2,\mathbb{R})_-\times U(1)_+$ global isometry. The explicit form of generators of $SL(2,\mathbb{R})_-$ are obtained from \eqref{sl2R-gen-GNC} upon the limit \eqref{near4} as
\be\label{sl2R-gen-GNC-NH}
\xi_1=\dfrac{\ell}{2}\p_{\hat u},\ \ \xi_2=\hat u\p_{\hat u} -\hat r \p_{\hat r} + \dfrac{\ell}{2 r_H(\hat{\phi})}\p_{\hat\phi},\ \  \xi_3=\dfrac{2 \hat u^2}{\ell}\p_{\hat u} + (\ell- \dfrac{4 \hat u \hat r}{\ell}) \partial_{\hat r}+\dfrac{2 \hat u}{r_H(\hat{\phi})}\p_{\hat \phi}.
\ee

Let us now analyze the presymplectic form  and the corresponding charges. To this end, we first recall that we obtained in section \ref{retarded-coord-sec} that
 both the Lee-Wald and the invariant presymplectic form  vanish on-shell for the general case. Therefore, both presymplectic structures also vanish for the special case $L_-=0$. All transformations that preserve the phase space are therefore either symplectic symmetries or pure gauge transformations, depending on whether or not they are associated with non-vanishing conserved charges.

{The symplectic symmetry vector field generators $\hat\xi$  may naively be defined from \eqref{retarded-symp-sym}, where we set $L_-=Y_-=0$ and  take the above near horizon limit. Doing so we obtain:}
$$
\hat{\xi}=(Y-\frac{\ell}{2r_H}Y')\p_{\hat{\phi}}-\frac{1}{\eps}\bigg(r_H Y-\frac{\ell Y'}{2}\bigg)'\p_{\hat r},
$$
where $Y=Y(\hat{\phi}), r_H=r_H(\hat{\phi})$ and primes denotes derivatives with respect to $\hat\phi$. Since this vector field admits a diverging $1/\eps$ term, it is not well-defined in the near-horizon limit. {Moreover, this vector field does not generate perturbations tangent to the near-horizon phase space. In doing the near-horizon change of coordinates, it is required to change the generator of symplectic symmetries.} One may check that a term like $f(\hat{\phi})\p_{\hat r}$ for both Barnich-Brandt or Iyer-Wald charges is pure gauge since it  does not contribute to the charges. Therefore, the problematic $1/\eps$ term may be dropped from $\hat{\xi}$ to obtain
\be\label{retarded-symp-sym-extremal}
\hat{\xi}=(Y-\frac{\ell}{2r_H}Y')\p_{\hat{\phi}}\,.
\ee
In fact, the vector field \eqref{retarded-symp-sym-extremal} is the correct vector field in the near-horizon phase space since $\mathcal L_\xi g_{\mu\nu}$ is tangent to the phase space \eqref{NBTZ4} with the transformation law
\bea
\delta_\xi r_H = r_H \p_{\hat\phi} Y + Y \p_{\hat\phi} r_H - \frac{\ell}{2}\p_{\hat\phi}^2 Y \, .\label{transrp}
\eea
This transformation law is consistent with the definition \eqref{horizon} and the Virasoro transformation law \eqref{Virl}. It is stricking that the resulting symplectic symmetry generator \eqref{retarded-symp-sym-extremal} takes a quite different form from \eqref{NH-chi-Banados} as well as all other ansatzes in the literature \cite{Balasubramanian:2009bg,Guica:2008mu,Compere:2014cna, Compere:2015mza}.

Using the expansion in modes $Y=e^{i n\hat\phi}$ we define the resulting vector field $l_n$. Since the vector field is field-dependent, we should use the ``adjusted bracket''  defined in section \ref{modified-bracket-sec}. Doing so, we obtain the Witt algebra
\bea
i[l_m,l_n]_{*}= (m-n) l_{m+n}.
\eea

One may then check that the surface charges associated with $\hat\xi$ are integrable, using the integrability condition for general field dependent generators, \emph{cf.} discussions of Appendix \ref{integrability-appendix-sec}. For the surface charges the Barnich-Brandt  and Iyer-Wald prescriptions totally agree since the invariant and Lee-Wald presymplectic forms coincide off-shell. We then obtain
\bea
Q_{\hat\xi}  = \frac{1}{8\pi G} \int \left( \frac{r_H^2}{\ell}Y - r_H Y'\right)  d\hat \phi.
\eea
After adding a boundary term $\boldsymbol d \boldsymbol B_{\hat\xi}$ where,
\bea
\boldsymbol B_{\hat\xi}   = \frac{1}{8\pi G} r_H Y,\label{bnd2}
\eea
to  the integrand and after using \eqref{horizon}, we find the standard Virasoro charge
\bea
Q_{\hat\xi} = \frac{\ell}{8\pi G} \int  L_+(\hat\phi) Y(\hat \phi)  d\hat \phi.
\eea
We have therefore shown that the near-horizon Virasoro symplectic symmetry can be directly mapped to the Brown-Henneaux {asymptotic symmetry at the boundary of AdS$_3$}.

\section{Discussion and outlook}\label{discussion-sec}

We established that the set of all locally AdS$_3$ geometries with Brown-Henneaux boundary conditions form a phase space whose total symmetry group is \emph{in general} a direct product between the left and right sector and between $U(1)$ Killing and Virasoro symplectic symmetries quotiented by a compact $U(1)$:
\bea
\left(U(1)_+ \times \frac{Vir_+}{U(1)_+}\right) \times \left(U(1)_- \times \frac{Vir_-}{U(1)_-}\right).\label{red1}
\eea
Elements of the phase space are solutions with two copies of ``Virasoro hair'' which can have two different natures: either Killing symmetry charges or symplectic symmetry charges. {One special patch of the phase space consists of the set of descendants of the global AdS$_3$ vacuum, where the two compact $U(1)$'s are replaced with two $SL(2,\mathbb R)$'s with compact $U(1)$ subgroup:
\bea
\left(SL(2,\mathbb{R})_+\times \frac{Vir_+}{SL(2,\mathbb R)_+}\right) \times \left(SL(2,\mathbb{R})_-\times \frac{Vir_-}{SL(2,\mathbb R)_-}\right).\label{red2}
\eea
In the case of the phase space with Poincar\'e AdS boundary conditions, the $U(1)$'s are instead non-compact. }

{In the case of the decoupling (near-horizon) limit of extremal black holes, the (let say) right sector is frozen to $L_-=0$ in order to be able to define the decoupling limit. In the limit the  $U(1)_-$ isometry is enhanced to $SL(2,\mathbb{R})_-$ and the $U(1)_-$ subgroup decompactifies. The exact symmetry group of the near-horizon phase space is a direct product of the left-moving Killing and left-moving non-trivial symplectic symmetries, isomorphic to Virasoro group quotiented by a compact $U(1)_+$,
\bea
SL(2,\mathbb{R})_- \times \left(U(1)_+ \times \frac{Vir_+}{U(1)_+}\right).
\eea
The global Killing $SL(2,\mathbb R)_-$ charges are fixed to zero, and there is no right-moving symplectic symmetry.}  We studied two particular decoupling limits which realized this symmetry. Taking the decoupling limit in Fefferman-Graham coordinates leads to zooming at fixed coordinate horizon radius while taking the decoupling limit in Gaussian null coordinates amounts to zooming on a wiggling horizon radius. We noticed that both decoupling limits lead to the same charge algebra.
In principle it should also be possible to have geometries associated with $$\left(SL(2,\mathbb{R})_+\times \frac{Vir_+}{SL(2,\mathbb R)_+}\right) \times \left(U(1)_-\times \frac{Vir_-}{U(1)_-}\right),$$ where the representative of the left-movers is fixed to have $L_+=-1/4$.

\paragraph{Orbits and Killing charges.} {The above obviously parallels the construction of Virasoro coadjoint orbits where the group that quotients the Virasoro group} is  the ``stabilizer group'' \cite{Witten:1987ty,Balog:1997zz}.  The stabilizer group, as intuitively expected, appears as the Killing isometry algebra of the locally AdS$_3$ geometries. Importantly for making connection with Virasoro orbits, the Killing vectors  commute with the Virasoro symmetries. Their associated conserved charges $J_\pm$ therefore label individual orbits. There are, nonetheless, other options for the stabilizer group besides compact $U(1)$ and $SL(2,\mathbb{R})$ which are, in general, labeled by $n$-fold cover of these stabilizer groups. This will lead to an extra integer label which being discrete, is not covered in the analysis of the type we presented here. This may be associated with a topological charge \cite{Sheikh-Jabbari:2015nn}.

\paragraph{Relationship with asymptotic symmetries.} It is well-known that all geometries with Brown-Henneaux boundary conditions admit two copies of the Virasoro group as asymptotic symmetry group \cite{Brown:1986nw}:
\bea
Vir_+ \times Vir_-\, .
\eea
In the case of the vacuum AdS$_3$ orbit, the global asymptotic $SL(2,\mathbb{R})_+ \times SL(2,\mathbb{R})_-$ subgroup of the Virasoro group exactly coincides with the
$SL(2,\mathbb{R})\times SL(2,\mathbb{R})$ isometries with  constant charges and the asymptotic symmetries reduce to \eqref{red2}. For generic orbits, only a $U(1)_+ \times U(1)_-$ subgroup of the $SL(2,\mathbb{R})_+ \times SL(2,\mathbb{R})_-$ is an isometry while the remaining generators are symplectic symmetries, which matches with \eqref{red1}.  The novelty is that the conserved charges are not defined at infinity only, they are defined at finite radius.

\paragraph{Symplectic charges and the Gauss law.} The electric charge of a set of electrons can be computing as the integral of the electric flux on an enclosing surface. It was observed some time ago that Killing symmetries lead to the same property for gravity \cite{2003CQGra..20.3685B}. The total mass of a set of isolated masses at equilibrium can be obtained by integrating the Killing surface charge on an enclosing surface. This property arises after viewing gravity as a gauge theory on the same footing as Maxwell theory. Here, we generalized the result of \cite{2003CQGra..20.3685B} to symplectic symmetries. Given a configuration with a symplectic symmetry and given a surface in a given homology class, one can define associated  symplectic charge which is conserved upon smoothly deforming the surface.

\paragraph{On the presymplectic form.} We reviewed the definition of the Lee-Wald and the invariant presymplectic forms and noticed that only the invariant one was  vanishing on-shell in both Fefferman-Graham and Gaussian null coordinates. This enabled us to define  symplectic symmetries on any closed circle which encloses all geometrical and topological defects. Together with the Killing symmetries, they extend the asymptotic symmetries of Brown-Henneaux in the bulk spacetime with identical results in both coordinate systems. However, the Lee-Wald presymplectic structure is equal on-shell to a boundary term in Fefferman-Graham coordinates. A natural question is whether a suitable boundary term can be added to the Lee-Wald presymplectic structure which fits among the known ambiguities in order that it vanishes exactly on-shell. We expect that it would be possible, but for our purposes the existence of an on-shell vanishing presymplectic structure was sufficient.

\paragraph{Coordinate independence and gauge transformations.} Every structure  we could find in Fefferman-Graham coordinates could be mapped onto the same structure in Gaussian null coordinates. We therefore expect that there is a gauge transformation between these coordinate systems which can be defined in the bulk spacetime. On general grounds, we expect that one could enhance the set of metrics with additional gauge transformation redundancy and incorporate more equivalent coordinate systems. Such a procedure would however not add  any physics classically since the physical phase space and charges would be left invariant. The advantage of either Fefferman-Graham or Gaussian null coordinates is that their only admissible coordinate transformations (which preserve the coordinates) are the physical symplectic and Killing symmetries. In that sense, they allow to express the phase space in a fixed gauge.

{
\paragraph{Generalization to other boundary conditions.} Boundary conditions alternative to Dirichlet boundary conditions exist for AdS$_3$ Einstein gravity \cite{Compere:2008us,Compere:2013bya,Troessaert:2013fma}. Our considerations directly apply to these boundary conditions as well. As an illustration, the semi-direct product of Virasoro and Ka{c}-Moody asymptotic symmetries found for chiral boundary conditions \cite{Compere:2013bya} can be extended to symplectic symmetries in the corresponding phase space. Indeed, it is easy to check that both the Lee-Wald and the invariant symplectic structures vanish for arbitrary elements in, and tangent to, the phase space. The BTZ black holes equipped with Virasoro and Ka{c}-Moody charges can be qualified as BTZ black holes with Virasoro and Ka{c}-Moody hair all the way to the horizon.
}

{
\paragraph{Generalization to Chern-Simons theories.} Three dimensional Chern-Simons theories are also theories without bulk dynamics. It is therefore natural to expect that any asymptotic symmetry will be a combination of Killing and symplectic symmetries. In the example of the $SL(2,\mathbb R) \times SL(2,\mathbb R)$ gauge group and Brown-Henneaux boundary conditions, it follows from the explicit definition of the phase space that the presymplectic form identically vanishes. Indeed, either using Fefferman-Graham or Null Gaussian coordinates, (3.11) or (3.19) of \cite{Compere:2014cna}, one directly gets that $\boldsymbol \omega \propto Tr(\delta_1 \boldsymbol A \wedge \delta_2 \boldsymbol A-\delta_2 \boldsymbol A \wedge \delta_1 \boldsymbol A)=0$ since $\delta \boldsymbol  A \propto dt^+$ in the left sector and $\delta \boldsymbol  A \propto dt^-$ in the right sector and both sectors are orthogonal in the trace. The result similarly follows for higher spin gauge theories. The asymptotic symmetries discussed in \cite{Henneaux:2010xg,Campoleoni:2010zq,Gaberdiel:2011wb,Campoleoni:2011hg,Ammon:2011nk,Compere:2013gja,Henneaux:2013dra} can therefore be promoted to a combination of Killing and symplectic symmetries. 
}

\section*{Acknowledgements}

We would like to thank Glenn Barnich, Kamal Hajian and especially Hossein Yavartanoo for fruitful discussions. AS would like to thank the Physics department at the Universit\'{e} Libre de Bruxelles and the International Solvay Institutes for hospitality during the start of this project. MMSHJ and AS  would like to thank Allameh Tabatabaii Prize Grant of Boniad Melli Nokhbegan of Iran and the ICTP network project NET-68.  G.C. is a Research Associate of the Fonds de la Recherche Scientifique F.R.S.-FNRS (Belgium) and he acknowledges the current support of the ERC Starting Grant 335146 ``HoloBHC" and the convention IISN 4.4503.15 of FNRS-Belgium. P.~Mao is supported by a PhD fellowship from the China
Scholarship Council.

\appendix
\section{Conserved charges for field dependent vectors}\label{appendix-A-sec}

In this appendix, we provide with the formalism of conserved charges in Einstein gravity in the case of field dependent vectors like those in \eqref{symplectic symmetries} and establish that the expression of charges obtained from covariant phase space methods \cite{Iyer:1994ys} or cohomological methods \cite{Barnich:2007bf} apply to this case as well. We also discuss the integrability of charge variations in the case of field dependent vectors. {We will keep the spacetime dimension arbitrary since no special feature arises in three dimensions. }

\subsection{Expression for the charges}

\paragraph{Field dependence and the Iyer-Wald charge.} Assume we have a vector $\chi$ which is a function of the dynamical fields $\Phi$ such as the metric. In our example, the metric dependence reduces to  $\chi=\chi(L_+,L_-)$. We call this a field dependent vector. We want to find the corresponding charge $\delta \boldsymbol{Q}_\chi$ and the integrability condition for such vectors.
We proceed using the approach of Iyer-Wald \cite{Iyer:1994ys} and carefully keep track of the field dependence. We adopt the convention that $\delta\Phi$ are Grassman even. First define the Noether current associated to the vector $\chi$ as
\begin{align}
\boldsymbol{J}_\chi[\Phi]&=\boldsymbol{\Theta}[\delta_\chi\Phi,\Phi]-\chi\cdot \boldsymbol{L}[\Phi],
\end{align}
where $\boldsymbol{L}[\Phi]$ is the Lagrangian (as a top form), and $\boldsymbol{\Theta}[\delta_\chi\Phi,\Phi]$ is equal to the boundary term in the variation of the Lagrangian, i.e $\de \boldsymbol{L}=\frac{\delta L}{\delta \Phi}\delta \Phi + d\boldsymbol{\Theta}[\delta\Phi,\Phi]$.  Using the Noether identities one can then define the on-shell vanishing Noether current as $\frac{\delta L}{\delta \Phi}\mathcal L_\chi \Phi = d \boldsymbol S_\chi[\Phi]$. It follows that $\boldsymbol J_\chi + \boldsymbol S_\chi$ is closed off-shell and therefore $\boldsymbol{J}_\chi\approx d \boldsymbol{Q}_\chi$, where $\boldsymbol{Q}_\chi$ is the Noether charge density (we use the symbol $\approx$ to denote an on-shell equality).
Now take a variation of the above equation
\begin{align}
\delta \boldsymbol{J}_\chi&=\delta \boldsymbol{\Theta}[\delta_\chi\Phi,\Phi]-\delta(\chi\cdot \boldsymbol{L})\cr
&=\delta \boldsymbol{\Theta}[\delta_\chi\Phi,\Phi]-\chi\cdot \delta \boldsymbol{L}-\delta\chi\cdot \boldsymbol{L} \cr
&\approx\delta \boldsymbol{\Theta}[\delta_\chi\Phi,\Phi]-\chi\cdot d\boldsymbol{\Theta}[\delta \Phi,\Phi]-\delta\chi\cdot\boldsymbol{L}\,.
\end{align}
Using the Cartan identity  $\mathcal{L}_\chi \boldsymbol{\sigma}=\chi\cdot d\boldsymbol{\sigma}+d(\chi\cdot \boldsymbol{\sigma})$ valid for any vector $\chi$ and any form $\boldsymbol{\sigma}$, we find
\begin{align}
\delta \boldsymbol{J}_\chi&=\Big(\delta \boldsymbol{\Theta}[\delta_\chi\Phi,\Phi]-\delta_\chi\boldsymbol{\Theta}[\delta \Phi,\Phi]\Big)+d(\chi\cdot\boldsymbol{\Theta}[\delta\Phi,\Phi])-\delta\chi\cdot \boldsymbol{L}.
\end{align}
The important point here is that
\begin{align}\label{delta vs deltaPhi}
\delta \boldsymbol{\Theta}[\delta_\chi\Phi,\Phi]&= \delta^{[\Phi]} \boldsymbol{\Theta}[\delta_\chi\Phi,\Phi]+ \boldsymbol{\Theta}[\delta_{\delta\chi}\Phi,\Phi]\,,
\end{align}
where we define $\delta^{[\Phi]}$ to act only on the explicit dependence on dynamical fields and its derivatives, but not  on the implicit field dependence in $\chi$. Therefore, we find
\begin{align}\label{presymplectic LW}
\nonumber \delta \boldsymbol{J}_\chi&=\Big(\delta^{[\Phi]} \boldsymbol{\Theta}[\delta_\chi\Phi,\Phi]-\delta_\chi\boldsymbol{\Theta}[\delta \Phi,\Phi]\Big)+d(\chi\cdot\boldsymbol{\Theta}[\delta\Phi,\Phi])+\Big(\boldsymbol{\Theta}[\delta_{\delta\chi}\Phi,\Phi]-\delta\chi\cdot \boldsymbol{L}\Big)\\
&=\boldsymbol{\omega}^{LW}[\delta\Phi\,,\,\delta_\chi\Phi\,;\,\Phi] + d(\chi\cdot\boldsymbol{\Theta}[\delta\Phi,\Phi])+\boldsymbol{J}_{\delta\chi}\,,
\end{align}
where
\be
\boldsymbol{\omega}^{LW}[\delta\Phi\,,\,\delta_\chi\Phi\,;\,\Phi]=\delta^{[\Phi]} \boldsymbol{\Theta}[\delta_\chi\Phi,\Phi]-\delta_\chi\boldsymbol{\Theta}[\delta \Phi,\Phi],
\ee
is the Lee-Wald presymplectic form \cite{Lee:1990nz}. Note that the variation acting on $\boldsymbol{\Theta}[\delta_\chi\Phi,\Phi]$, only acts on the explicit field dependence. This is necessary in order for $\boldsymbol{\omega}^{LW}[\delta\Phi\,,\,\delta_\chi\Phi\,;\,\Phi]$ to be bilinear in its variations. Reordering the terms we find
\begin{align}
\nonumber \boldsymbol{\omega}^{LW}[\delta\Phi\,,\,\delta_\chi\Phi\,;\,\Phi]&=\delta \boldsymbol{J}_\chi-\boldsymbol{J}_{\delta\chi}-d(\chi\cdot\boldsymbol{\Theta}[\delta\Phi,\Phi])\\
&=\delta^{[\Phi]}\boldsymbol{J}_\chi -d(\chi\cdot\boldsymbol{\Theta}[\delta\Phi,\Phi]).
\end{align}

If $\delta \Phi$ solves  the linearized field equations, then $\boldsymbol{J}_\chi\approx d \boldsymbol{Q}_\chi$ implies  $\delta^{[\Phi]}\boldsymbol{J}_\chi \approx d(\delta^{[\Phi]}\boldsymbol{Q}_\chi)$. As a result we obtain
\bea\label{omega vs charge}
\boldsymbol{\omega}^{LW}[\delta\Phi\,,\,\delta_\chi\Phi\,;\,\Phi]& \approx d \boldsymbol{k}_\chi^{IW} [\delta\Phi; \Phi]
\eea
where $ \boldsymbol{k}_\chi^{IW}$ is the Iyer-Wald surface charge form
\bea
 \boldsymbol{k}_\chi^{IW} = \Big(\delta^{[\Phi]}\boldsymbol{Q}_\chi-\chi\cdot\boldsymbol{\Theta}[\delta\Phi,\Phi]\Big).
\eea
Therefore the infinitesimal charge associated to a field dependent vector and a codimension two, spacelike compact surface $S$ is defined as the Iyer-Wald charge
\begin{align}\label{charge variation}
\delta H_\chi &= \oint_S \boldsymbol{k}_\chi^{IW} [\delta\Phi; \Phi]= \oint_S \Big(\delta^{[\Phi]}\boldsymbol{Q}_\chi-\chi\cdot\boldsymbol{\Theta}[\delta\Phi,\Phi]\Big).
\end{align}
The key point in the above expression is that the variation does not act on $\chi$. {One may rewrite the charge as
\begin{align}\label{charge variation2}
\delta H_\chi &= \oint_S \Big(\delta \boldsymbol{Q}_\chi -\boldsymbol{Q}_{\delta \chi}  -\chi\cdot\boldsymbol{\Theta}[\delta\Phi,\Phi]\Big).
\end{align}
From the above, there is an additional term in the Iyer-Wald charge in the case of field dependent vectors.}

 \paragraph{Field dependence and the Barnich-Brandt charge.} There is another definition of the presymplectic structure which leads to a consistent covariant phase space framework. This is the so-called invariant presymplectic form \cite{Barnich:2007bf} defined through Anderson's homotopy operator \cite{Andersonbook}:
\bea
\bomega^{inv}[\delta_1 \Phi ,\delta_2 \Phi;\Phi ]  = -\frac{1}{2} I_{\delta_1 \Phi}^n \left( \delta_2 \Phi^i \frac{\delta \bL}{\delta \Phi^i} \right) - (1 \leftrightarrow 2) ,\\
I^n_{\delta \Phi}  \equiv \left( \delta \Phi^i \frac{\p}{\Phi_{\, ,\mu}^i} - \delta \Phi^i \p_\nu \frac{\p}{\Phi_{\, ,\nu\mu}^i} + \delta\Phi_{,\nu} \frac{\p}{\p \Phi_{,\nu\mu}}\right) \frac{\p}{\p d x^\mu} .\nn
\eea
The invariant presymplectic form only depends on the equations of motion of the Lagrangian and is therefore independent on the addition of boundary terms in the action. This presymplectic structure differs from the Lee-Wald presymplectic structure by a specific boundary term $\bE$
\begin{equation}
\bomega^{inv}[\delta_1 \Phi ,\delta_2 \Phi;\Phi ]  =\bomega^{LW}[\delta_1 \Phi ,\delta_2 \Phi;\Phi ]  + { d} \bE[\delta_1 \Phi ,\delta_2 \Phi,\Phi ] ,\label{id1}
\end{equation}
where $\bE$ is given by \cite{Barnich:2007bf,Compere:2007az}
\begin{equation}
\bE[\delta_1 \Phi ,\delta_2 \Phi,\Phi ]  = -\half I^{n-1}_{\delta_1\Phi} \bTheta[\delta_2 \Phi , \Phi ] - (1 \leftrightarrow 2).
\end{equation}
Here, $\bTheta[\delta \Phi , \Phi]$ is defined as $I^n_{\delta \Phi}  \boldsymbol{L}$, which agrees with the Lee-Wald prescription and Anderson's homotopy operator for a $n-1$ form is given for second order theories by
\bea
I_{\delta \Phi}^{n-1} &\equiv & \left(  \frac12\delta\Phi^i \frac\partial{\partial\Phi^i_{,\nu }}
-\frac13\delta\Phi^i \partial_\rho  \frac\partial{\partial\Phi^i_{,\rho \nu }}
+\frac23\delta\Phi^i_{,\rho } \frac\partial{\partial\Phi^i_{, \rho \nu}} \right) \frac{\p}{\p d x^\nu} .
\eea
The identity \eqref{id1} follows from $[\delta,I_{\delta \Phi}^{p}]=0$ and the equalities
\bea\label{comm1}
0 \leqslant p < n:\quad I_{\delta \Phi}^{p+1} d + d I_{\delta \Phi}^{p} = \delta,\\
p=n:\quad \delta \Phi^i \frac{\delta }{\delta \Phi^i} + d I_{\delta \Phi}^{n} = \delta .
\eea

The  presymplectic structure evaluated on the field transformation generated by the (possibly field-dependent) vector field $\chi$,  $\bomega^{inv}[\delta_1 \Phi ,\delta_\chi \Phi;\Phi ]$, is defined from a contraction as
\bea
\bomega^{inv}[\delta_1 \Phi ,\delta_\chi \Phi;\Phi ]=(\p_{(\mu)}\delta_\chi \Phi)\dfrac{\p}{ \delta_2 \Phi^i_{(\mu)}}\bomega^{inv}[\delta_1 \Phi ,\delta_2 \Phi;\Phi ].
\eea
It then follows from \eqref{id1} that
\bea
\bomega^{inv}[\delta \Phi ,\delta_\chi \Phi;\Phi ]=\boldsymbol{\omega}^{LW}[\delta\Phi\,,\,\delta_\chi\Phi\,;\,\Phi] + d \bE[\delta \Phi,\delta_\chi \Phi,\Phi].
\eea
Inserting \eqref{omega vs charge} from the above analysis, we find
\begin{align}\label{BB omega vs charge}
\boldsymbol{\omega}^{inv}[\delta\Phi\,,\,\delta_\chi\Phi\,;\,\Phi]& \approx d \boldsymbol{k}_\chi^{BB} [\delta\Phi; \Phi] 
\end{align}
where $ \boldsymbol{k}_\chi^{BB}$ is the Barnich-Brandt surface charge form, 
\begin{align}\label{BBch}
 \boldsymbol{k}_\chi^{BB} [\delta\Phi; \Phi]  = \delta^{[\Phi]}\boldsymbol{Q}_\chi -\chi\cdot\boldsymbol{\Theta}[\delta\Phi,\Phi]+ \bE[\delta \Phi,\delta_\chi \Phi,\Phi]
.
\end{align}
After evaluation on a codimension two, spacelike compact surface  $S$, the infinitesimal charge is
\begin{align}\label{BB charge}
\delta H_\chi & \equiv \oint_S \boldsymbol{k}_\chi^{BB} [\delta\Phi; \Phi]= \oint_S \Big(\delta^{[\Phi]}\boldsymbol{Q}_\chi -\chi\cdot\boldsymbol{\Theta}[\delta\Phi,\Phi]+ \bE[\delta \Phi,\delta_\chi \Phi,\Phi]
\Big).
\end{align}
This formula is identical to the standard Barnich-Brandt formula, which is therefore valid even when $\chi$ has an implicit field dependence.

The Barnich-Brandt surface charge form can be alternatively defined as $\boldsymbol{k}_\chi^{BB} [\delta\Phi; \Phi] = I_{\delta\Phi}^{n-1}\boldsymbol S_\chi [\Phi]$ where $\boldsymbol S_\chi $ is the on-shell vanishing Noether current defined earlier. Here the formalism requires that the homotopy operator only acts on the explicit field dependence in $\boldsymbol S_\chi [\Phi]$ but not on the possible implicit field dependence in $\chi$. Otherwise the commutation relations \eqref{comm1} would not be obeyed. (Also, if the operator $ I_{\delta\Phi}^{n-1}$ acts anyways on the field-dependence in $\chi$, the resulting terms will vanish on-shell by definition of $\boldsymbol S_\chi [\Phi]$.)
One can then show that this definition is equivalent on-shell to $\boldsymbol{k}_\chi^{BB} [\delta\Phi; \Phi] = I_\chi^{n-1}  \boldsymbol{\omega}^{inv}[\delta\Phi\,,\,\delta_\chi\Phi\,;\,\Phi]$ where the homotopy operator $I_\chi^{n-1}$ obeys $d I_\chi^{n-2}  +I_\chi^{n-1}  d=1$ \cite{Barnich:2007bf,Compere:2007az}. For the purposes of this homotopy operator, $\chi$ is considered as a field by itself and the implicit field dependence in $\Phi$ is irrelevant. One always obtains the same expression \eqref{BBch}.

A special feature of the cohomological formalism is that the presymplectic form is not identically closed in the sense that
\bea\nn
\delta^{[\Phi]}_1 \bomega^{inv} [\delta_2 \Phi,\delta_3 \Phi , \Phi] + (2,3,1)+(3,1,2) = d \big[ \delta^{[\Phi]}_1 \bE[\delta_2 \Phi,\delta_3\Phi,\Phi] +  (2,3,1)+(3,1,2)  \big]
\eea
is a boundary term, not zero. A prerequisite in order to have a well-defined charge algebra is that in the phase space
\bea\label{cocycle}
\oint \left( \delta^{[\Phi]}_1 \bE[\delta_2 \Phi,\delta_\chi\Phi,\Phi] + \delta^{[\Phi]}_2 \bE[\delta_\chi \Phi,\delta_1\Phi,\Phi] +\delta_\chi \bE[\delta_1 \Phi,\delta_2\Phi,\Phi]  \right) =0.
\eea
This condition will be obeyed for the phase spaces considered here.

In $3d$ Einstein theory, the charges are given explicitly by
\begin{align}\label{kgrav}
\mathbf{k}_\chi^{Einstein}
&= \dfrac{\sqrt{-g}}{8\pi G}
(d^{n-2}x)_{\mu\nu}
\bigg\{
\chi^\nu\nabla^\mu h
-\chi^\nu\nabla_\sigma h^{\mu\sigma}
+\chi_\sigma\nabla^{\nu}h^{\mu\sigma}
+\frac{1}{2}h\nabla^{\nu} \chi^{\mu}
-h^{\rho\nu}\nabla_\rho\chi^{\mu}\nonumber\\
&\hspace{9cm}
+ \frac{\alpha}{2}
h^{\sigma\nu}(\nabla^\mu\chi_\sigma +
\nabla_\sigma\chi^\mu)
\bigg\}\,,
\end{align}
where $\alpha=0$ according to the definition of Iyer-Wald and $\alpha=+1$ according to the definition by Barnich-Brandt. Here $(d^{n-2}x)_{\mu\nu} = \frac{1}{2}\eps_{\mu\nu\alpha}dx^\alpha$ in $3$ dimensions. The last prescription also coincides with the one of Abbott-Deser \cite{Abbott:1981ff}. In the case of Killing symmetries, there is no difference between the Iyer-Wald and Barnich-Brandt or Abbott-Deser charges. However, there is a potential difference for symplectic symmetries.

Equations \eqref{omega vs charge} or \eqref{BB omega vs charge} relates the charges computed on different surfaces. Consider the infinitesimal charges \eqref{charge variation} or \eqref{BB charge} evaluated on two different codimension two, spacelike compact surface  $\mathcal{S}_1$ and $\mathcal{S}_2$. Denote a surface joining these two  by $\Sigma$. Then taking the integral of \eqref{omega vs charge} or \eqref{BB omega vs charge} over $\Sigma$ and using Stokes' theorem, one obtains
\begin{align}\label{charge difference}
\delta H_\chi\Big\vert_{\mathcal{S}_2}-\delta H_\chi\Big\vert_{\mathcal{S}_1}&=\int_{\Sigma}\boldsymbol{\omega}[\delta\Phi\,,\,\delta_\chi\Phi\,;\,\Phi].
\end{align}
Killing symmetries ($\delta_\chi\Phi \approx 0$) or symplectic symmetries ($\boldsymbol{\omega}[\delta\Phi\,,\,\delta_\chi\Phi\,;\,\Phi] \approx 0 $, $\delta_\chi\Phi \neq 0$) therefore lead to conserved charges.

\subsection{Integrability of charges}\label{integrability-appendix-sec}

In order the charge perturbation defined in \eqref{charge variation} or \eqref{BB charge} to be the variation of a finite charge $H_\chi[\Phi]$ defined over any field configuration $\Phi$ connected to the reference configuration $\bar\Phi$ in the phase space, it should satisfy integrability conditions. More precisely, integrability implies that the charges defined as $H_\chi=\int_{\bar{\Phi}}^{\Phi} \delta H_\chi$ along a path $\gamma$ over the phase space does not depend upon $\gamma$. In the absence of topological obstructions in the phase space, it amounts to the following integrability conditions
\begin{align}
I\equiv\delta_1\delta_2H_\chi-\delta_2\delta_1H_\chi&=0.
\end{align}
which can be conveniently written as
\begin{align}
I\equiv \delta^{[\Phi]}_1\delta_2H_\chi + \delta_2 H_{\delta_1 \chi} - (1 \leftrightarrow 2)&=0.
\end{align}

Using \eqref{BB charge} in the first term we note that the Noether charge term drops by anti-symmetry in $(1 \leftrightarrow 2)$. We obtain
\bea
I &=&\oint \left(  \delta_1^{[\Phi]} \boldsymbol E[\delta_2 \Phi, \delta_\chi \Phi ; \Phi ] -\chi \cdot  \delta_1^{[\Phi]}  \boldsymbol \Theta[\delta_2 \Phi ; \Phi] \right)  + \delta_2 H_{\delta_1 \chi} - (1 \leftrightarrow 2).
\eea
We can then use the cocyle condition \eqref{cocycle} to obtain
\bea
I &=&\oint \left(  -\delta_\chi \boldsymbol E[\delta_1 \Phi, \delta_2 \Phi ; \Phi ] -\chi \cdot  \delta_1^{[\Phi]} \boldsymbol \Theta[\delta_2 \Phi ; \Phi] +\chi \cdot  \delta_2^{[\Phi]}  \boldsymbol\Theta[\delta_1 \Phi ; \Phi]  \right)  + \delta_2 H_{\delta_1 \chi} - \delta_1 H_{\delta_2 \chi} .\nn
\eea
We can replace $\delta_\chi $ by $\delta_\chi^{\Phi}$ or $\mathcal L_\chi$ in the first term. With the help of Cartan identity $\mathcal{L}_\chi  = d \chi \cdot + \chi \cdot d$ and using the definition of the invariant presymplectic form \eqref{id1} we finally obtain
\bea\label{field-dependent-integrability}
I=-\oint \chi\cdot \bomega^{inv}[\delta_1 \Phi ,\delta_2 \Phi;\Phi ]-\Big(\delta_1 H_{\delta_2\chi}-\delta_2 H_{\delta_1\chi}\Big)=0\,.
\eea
The term in parentheses arises due to the field dependence of vectors.

By dropping the $\bE$ term, one obtains the integrability condition for field dependent vectors according to the definition of charges of Iyer-Wald. The result is simply
\bea\label{field-dependent-integrability-LW}
I=-\oint \chi\cdot \bomega^{LW}[\delta_1 \Phi ,\delta_2 \Phi;\Phi ]-\Big(\delta_1 H_{\delta_2\chi}-\delta_2 H_{\delta_1\chi}\Big)=0\, .
\eea


\providecommand{\href}[2]{#2}\begingroup\raggedright\endgroup

\end{document}